\documentclass[preprints,article]{Definitions/mdpi} 
\firstpage{1} 
\pubvolume{9}
\issuenum{1}
\articlenumber{39}
\pubyear{2023}
\copyrightyear{2023}
\externaleditor{Academic Editors: Efe Yazgan, Pedro Ferreira da Silva and Santiago Peris}  
\datereceived{23 October 2022} 
\daterevised{19 December 2022} 
\dateaccepted{20 December 2022} 
\datepublished{6 January 2023} 
\hreflink{https://doi.org/\linebreak10.3390/universe9010039}
\pdfoutput=1

\Title{Recent Cross-Section Measurements of Top-Quark Pair Production in Association with Gauge Bosons}

\TitleCitation{Recent Cross-Section Measurements of Top-Quark Pair Production in Association with Gauge Boson}

\Author{Joshuha Thomas-Wilsker
  \orcidA{}}

\AuthorNames{Joshuha Thomas-Wilsker}

\AuthorCitation{Thomas-Wilsker, J.}

\address[1]{{Institute of High Energy Physics, Chinese Academy of Sciences, Beijing 100049, China}; jthomasw@cern.ch}

\abstract{
This article reviews recent cross-section measurements of $\text{t}\bar{\text{t}}$ production in association with a photon, W or Z boson at the Large Hadron Collider (LHC). All measurements reviewed use proton–proton (pp) datasets collected by the ATLAS and CMS experiments between 2016 and 2018 from collisions at a centre-of-mass energy of 13\,TeV during the LHC Run 2. Differential and inclusive cross-section measurements are discussed along with the constraints on the effective field theory operators accessible through each process. Finally, we discuss the potential for measurements of these processes at future colliders.
}


\begin{document}

\section{Introduction}
The top quark has several unique features that distinguish it from other Standard Model (SM) particles. With its electroweak (EW) scale mass of approximately $172$\,GeV it is by far the most massive of the fundamental SM particles. This mass, along with an associated Yukawa coupling value close to unity, suggests it may have a special role in the EW symmetry-breaking mechanism. It also has a uniquely short lifetime of ~$\mathcal{O}(10^{-25})$\,seconds which prevents it from hadronising before it decays\endnote{There have in fact been phenomenological investigations into the possible existence of top-quark pair-bound states, also known as `Toponium'~\cite{Toponium}, in order to explain excesses seen in LHC Run 2 data measurements of top-quark pair production with dilepton decays, where the top-quark pair is produced near threshold}, making it the only quark for which it is possible to study bare quark properties via its decay products.

This unconventional particle provides us with a tool with which we can scrutinise predictions of SM parameters and test a plethora of Beyond the Standard Model (BSM) hypotheses. Several model-dependent searches for BSM physics look for deviations in top-pair production rates and could verify theoretical models that predict the existence of top super-partners, vector-like quarks or even Dark Matter. There are also many model-independent searches that use an effective field theory framework to search for anomalous couplings. Additionally, there are many measurements at the LHC for which SM top-quark processes are important backgrounds and therefore also benefit from improved measurements in the top sector.

In proton–proton collisions at the LHC, the dominant top-quark production mechanism produces top quarks in pairs via the QCD process $\mathrm{gg \rightarrow \text{t}\bar{\text{t}}}$. Due to the CKM matrix element $|V_{tb}|$ being so large, the top-quark decays almost exclusively via the process $\mathrm{t\rightarrow bW}$. Thus, most top-quark pairs are produced via the interaction $\mathrm{gg \rightarrow \text{t}\bar{\text{t}} \rightarrow bW^+bW^-}$. The $\text{t}\bar{\text{t}}$ process is often categorised according to the decay of the two W bosons. These categories are referred to as dileptonic, semi-leptonic or full hadronic, and are often studied independently due to the varying backgrounds and final state signatures. 

\begin{figure}[H]
\centering
\captionsetup{justification=centering}
\includegraphics[width=5cm,height=4cm]{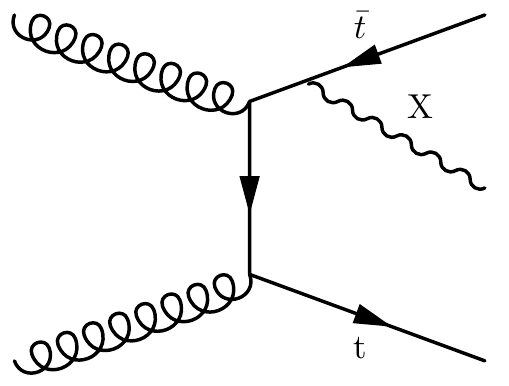}
\caption{Leading-order Feynman diagram for gluon-gluon top-pair production ($\mathrm{gg \rightarrow \text{t}\bar{\text{t}}}$) in association with a boson (X).\label{fig_ttX_feyn}}
\end{figure}

\textls[-5]{The focus of this article is on $\mathrm{\text{t}\bar{\text{t}}}$ production in association with an additional gauge boson ($\mathrm{\text{t}\bar{\text{t}}X}$), as exemplified in Figure \ref{fig_ttX_feyn}. More explicitly, the latest ATLAS and CMS cross-section measurements of $\mathrm{\text{t}\bar{\text{t}}}$ production in association with either a photon ($\gamma$), W or Z boson. These measurements typically assume SM-like processes to obtain inclusive and differential cross-sections; however, several of them also provide interpretations using the Standard Model Effective Field Theory (SMEFT) framework~\cite{SMEFT1,SMEFT2}. These processes provide a deep insight into the nature of the couplings in the top-quark interactions with gauge boson. The publications discussed focus on cross-section measurements performed using datasets collected during the LHC Run 2, where high energy ($13$\,TeV) collisions and huge datasets (approximately 140 fb$^{-1}$ integrated luminosity) make it possible investigate these rare $\text{t}\bar{\text{t}}$ processes in more detail than ever before. The future of these measurements is also discussed, focusing on their potential at the HL-LHC and the main future collider~candidates.}

\section{\boldmath{$\mathrm{\text{t}\bar{\text{t}}Z}$} Measurements}\label{Section:ttZ}
Inclusive and differential measurements of the $\mathrm{\text{t}\bar{\text{t}}Z}$ cross-section are interesting because they directly probe the coupling between the top quark and the neutral EW Z boson, also known as the t-Z coupling. Furthermore, several BSM theories~\cite{future_EW_couplings,precision_EW_Zres} also predict anomalous neutral EW top-quark couplings that can drastically change the amplitude and subsequently the measured cross-section. Such couplings have also been interpreted using an effective field theory (EFT) approach~\cite{ttZ_EFT}. The attraction here is that the EFT approach provides a model-independent way to interpret possible deviations in a cross-section measurement from its SM value.

This process is also an important background for several SM measurements, for example single-top production in association with a Z boson, $\mathrm{\text{t}\bar{\text{t}}H}$ and many BSM searches~\cite{SUSYOSSF}. A precise measurement of the process is therefore beneficial to analyses looking to minimise the uncertainties associated with this process.

The first measurements of $\mathrm{\text{t}\bar{\text{t}}Z}$ were performed by ATLAS and CMS at $8\,$TeV. However, both ATLAS and CMS have also measured this process using partial Run 2 datasets of 36.1~$\mathrm{fb^{-1}}$ and $35.9~\mathrm{fb^{-1}}$, respectively, from  13\,TeV  collisions, where the production rate increases approximately by a factor of 4~\cite{ATLASttZ2019,CMSttZ2018}.

Events were selected with two or more leptons to simultaneously extract the $\mathrm{\text{t}\bar{\text{t}}Z}$ and $\mathrm{\text{t}\bar{\text{t}}W}$ production cross-sections. The 3 and 4 lepton categories are the most sensitive to the $\mathrm{\text{t}\bar{\text{t}}Z}$ process. Observed and expected significance values in both experiments for the $\mathrm{\text{t}\bar{\text{t}}Z}$ process are well above 5$\sigma$ in these measurements. ATLAS measured the cross-section to be $\sigma_{\mathrm{\text{t}\bar{\text{t}}Z}} = 0.95 \pm 0.08(\text{stat}) \pm 0.10(\text{syst})$\,pb while CMS measured a value of $\sigma_{\mathrm{\text{t}\bar{\text{t}}Z}} = 0.99^{+0.09}_{-0.08}(\text{stat}) \pm ^{0.12}_{0.10}(\text{syst})$\,pb. One can see that, due to the large dataset, the statistical uncertainty is dramatically reduced and the systematic uncertainty on this result is already of a similar size. CMS also provide the first limits on anomalous t-Z couplings with $\mathrm{\text{t}\bar{\text{t}}Z}$ data using an effective field theory (EFT) framework. Typically, this process provides the tightest constraints on this coupling.

Both collaborations~\cite{CMSttZ2020,ATLASttZ2021} now measure $\mathrm{\text{t}\bar{\text{t}}Z}$ separately from $\mathrm{\text{t}\bar{\text{t}}W}$ using Run 2 datasets of $139~\mathrm{fb^{-1}}$ and $77.5~\mathrm{fb^{-1}}$ for ATLAS and CMS, respectively. In both analyses, events with 3 or 4 isolated leptons (electrons or muons) are selected, targeting processes where one or both top quarks decay leptonically along with leptonic decays of the Z boson. Event and object quality requirements ensure the leptons are isolated and consistent with either the decay of a W boson (from the top-quark decay) or a Z boson. B-tagging algorithms are used to distinguish jets that originate from the hadronisation of b-quarks from those originating from light (up, down , strange or charm) quarks or gluons. Events are then further categorised according to the flavour and multiplicity of the jets in the event.

\begin{figure}[H]
  \centering
  \captionsetup{justification=centering}
  \includegraphics[width=11cm,height=5.0cm]{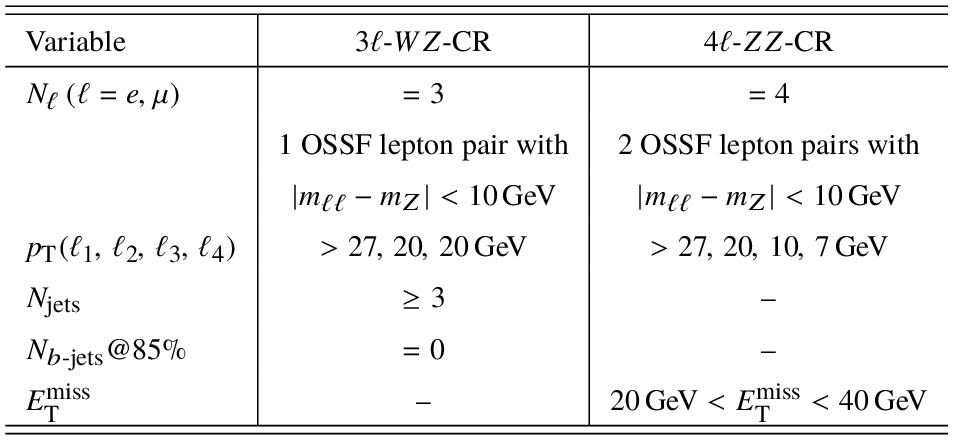}
  \caption{ATLAS control regions~\cite{ATLASttZ2021}. \label{ATLAS_ttZ_CR}}
\end{figure}

The ATLAS analysis selects events at detector level (using objects reconstructed from detector signals) with a minimum of two jets along with the aforementioned 3 or 4 lepton signature. Further signal region requirements are applied to maximise the sensitivity to $\mathrm{\text{t}\bar{\text{t}}Z}$ production while ensuring enough signal events are retained to prevent the statistical uncertainty from becoming too large in the differential measurement. Additionally, control regions are defined to estimate background contributions from processes with prompt leptons from EW boson decays. Control region definitions can be found in Figure \ref{ATLAS_ttZ_CR} where WZ/ZZ plus light jet processes dominate. The event yields from control regions are constrained by the observed data yields in these regions, which are then extrapolated to predict their contribution in the signal regions. WZ/ZZ plus b-jet production is not included in this method and are instead predicted directly using simulated templates which are included in the signal extraction procedure. 

Another significant background contribution comes from processes where the selected lepton is not from the prompt decay of a vector boson (aka non-prompt/fake-lepton). This background mostly stems from $\mathrm{\text{t}\bar{\text{t}}}$ dilepton processes where additional non-prompt leptons can originate from leptonically decaying heavy-flavour hadrons and/or jets that `fake' a leptonic signature and is subsequently misidentified as a lepton. The contribution from this background is estimated using the matrix-method ~\cite{MM1,MM2} which relies on the different probabilities that prompt and fake leptons pass the identification, isolation and impact parameter requirements. All other background processes are estimated from simulation, normalised to the latest theoretical cross-section prediction~\cite{ttZ_theoryXS_ATLAS_1,ttZ_theoryXS_2,ttZ_theoryXS_3}.

In comparison, the latest CMS inclusive cross-section measurement employs a very similar detector-level event selection. The measurement selects events with 3 or 4 lepton signatures and at least one jet. Events are then categorised according to the number of leptons, light (up, down, strange and gluon) flavour jets and heavy (bottom) flavour jets. The background processes are the same and are grouped in a mostly identical manner. All background processes with prompt leptons are modelled using the state-of-the-art simulation and normalised to the latest cross-section calculation. The normalisation of the WZ/ZZ plus jets processes are not extracted in the fit but are assigned uncertainties to cover the difference between data and the simulation in a dedicated control region. Backgrounds with fake/non-prompt leptons are estimated using the ``fake rate'' method in which estimates are made of the rate at which fake leptons pass the lepton selection requirements in control regions, and then this is extrapolated to the signal regions.

\begin{figure}[H]
\centering
\captionsetup{justification=centering}
\includegraphics[width=11cm,height=5.5cm]{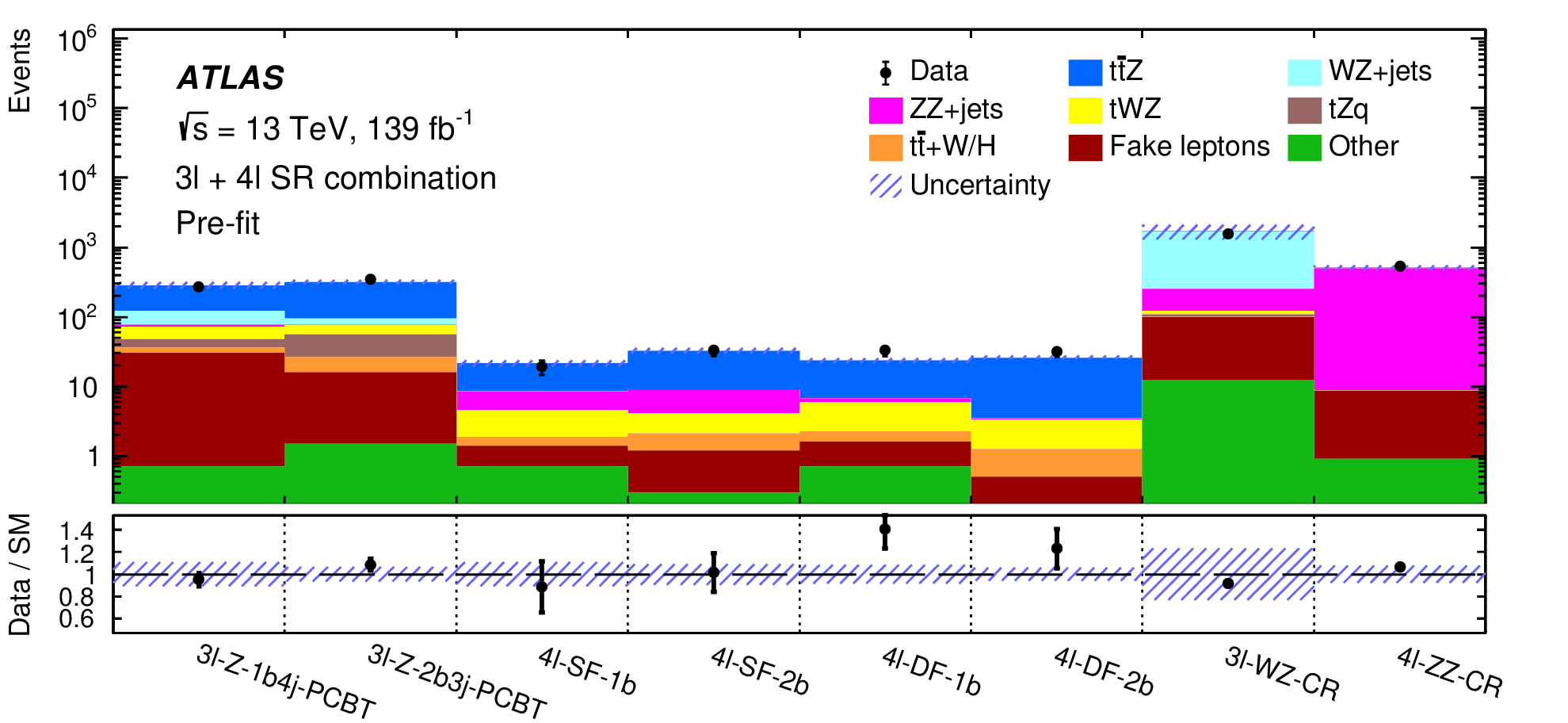}
\caption{ATLAS signal regions~\cite{ATLASttZ2021}. \label{ATLAS_ttZ_SR}}
\end{figure}

\begin{figure}[H]
\centering
\captionsetup{justification=centering}
\includegraphics[width=10cm,height=6cm]{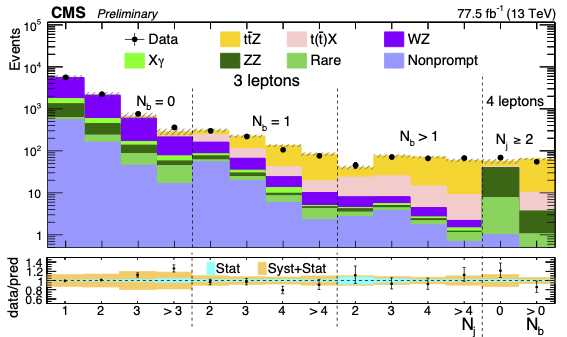}
\caption{CMS signal regions~\cite{CMSttZ2020}. \label{CMS_ttZ_SR}}
\end{figure}

Both analyses extract the inclusive cross-section through a simultaneous maximum likelihood fit of the predicted yields of the signal and background processes to data in the signal regions. The signal strength ($\mu = \frac{\sigma^{best\,fit}}{\sigma^{SM}}$) is a free parameter in the fit and uncertainties are included in the fit as nuisance parameters constrained by Gaussian functions. The ATLAS analysis simultaneously fits data in the control regions and the WZ/ZZ plus light jets backgrounds treated as free parameters in the fit. The yields for the fitted simulation and data in the signal regions for both analyses can been seen in \mbox{Figure{s} \ref{ATLAS_ttZ_SR} and \ref{CMS_ttZ_SR}}.  

The inclusive cross-section measured by ATLAS~\cite{ATLASttZ2021} from the combined fit in the 3 and 4 lepton signal regions, corresponding to a fiducial volume in which the Z-boson invariant mass lies between 70 and 110\,GeV, is found to be

\begin{equation}
    \sigma^{\text{pp}\rightarrow \text{t}\bar{\text{t}}Z}_{\mathrm{ATLAS}} = 0.99 \pm 0.05 \mathrm{(stat.)} \pm 0.08 \mathrm{(syst.)} \, \mathrm{pb}
\end{equation}
where the dominant systematic uncertainties originate from the $\mathrm{\text{t}\bar{\text{t}}Z}$ parton shower modelling, $\mathrm{tWZ}$ background modelling and the identification. 

The CMS cross-section measurement~\cite{CMSttZ2020} yielded a value of

\begin{equation}
    \sigma^{\text{pp}\rightarrow \text{t}\bar{\text{t}}Z}_{\mathrm{CMS}} = 1.00 ^{+0.06}_{-0.05} \mathrm{(stat.)} \pm ^{+0.07}_{-0.06} \mathrm{(syst.)} \, \mathrm{pb}
\end{equation}

The results are evidently in excellent agreement with one another and reasonable agreement with the SM theoretical prediction~\cite{ttZ_theoryXS_2,ttZ_theoryXS_3} of

\begin{equation}
    \sigma^{\text{pp}\rightarrow \text{t}\bar{\text{t}}Z}_{\mathrm{theo.}} = 0.88 ^{+0.09}_{-0.15} \, \mathrm{pb}
\end{equation}

Several differential cross-section measurements investigate the kinematics of the $\mathrm{\text{t}\bar{\text{t}}Z}$ system. In general, these measurements are performed by first subtracting background estimates from the data and then implementing an unfolding procedure which removes detector effects from the data so it can be compared with theoretical predictions. Migration matrices are constructed as part of the method that ensures resolution and acceptance affects are accounted for. The ATLAS measurement uses an iterative Bayesian unfolding to distributions defined using either particle or parton-level objects. Particle-level objects are defined using the collection of stable particles from the full matrix element plus parton shower simulation, i.e., baryons and mesons. Parton-level objects are defined using the unstable particles before any hadronisation effects have been simulated, i.e., quarks and~gluons. 

\begin{figure}[H]
\centering
\captionsetup{justification=centering}
\includegraphics[width=6.5cm,height=6.0cm]{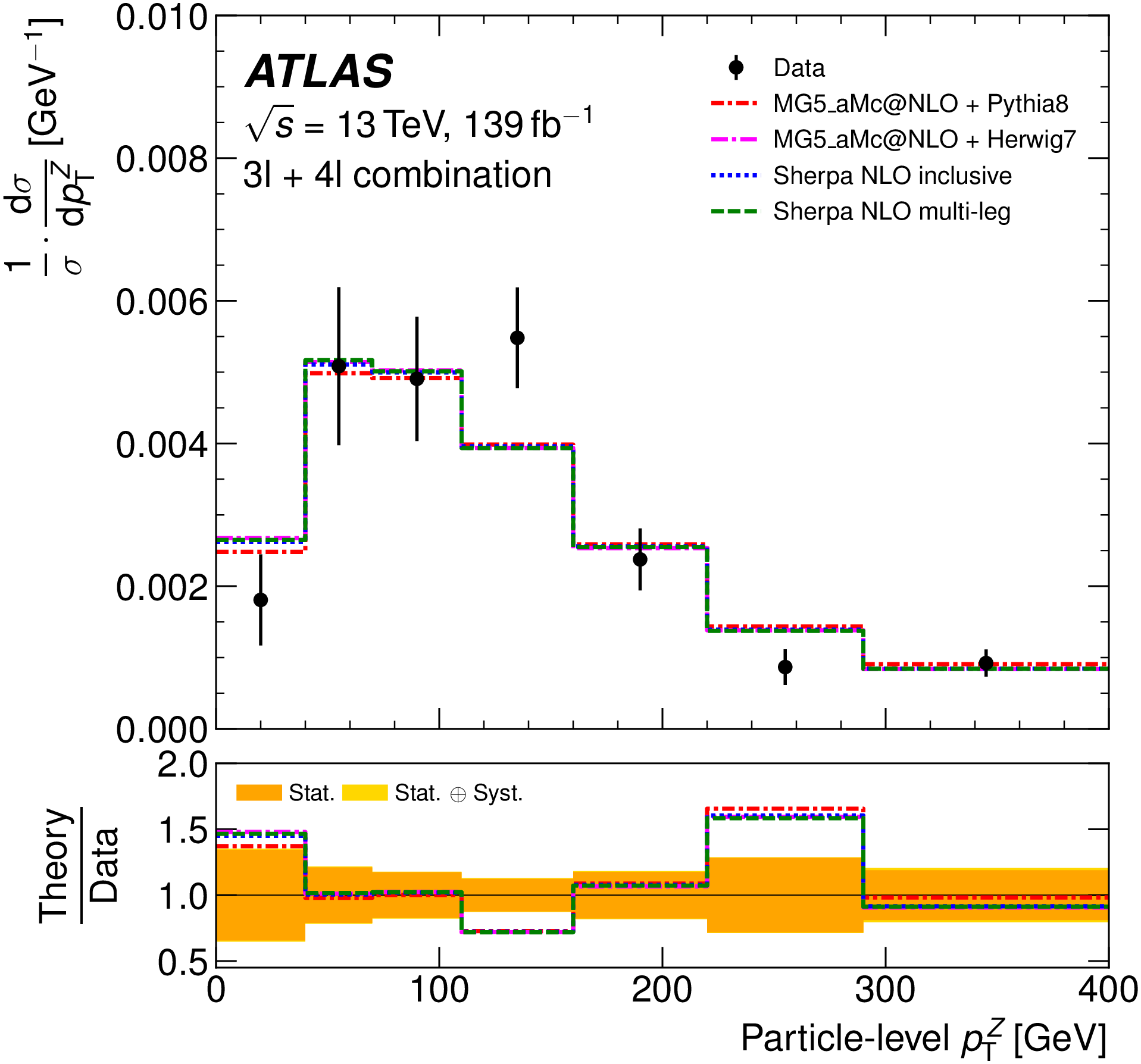}
\includegraphics[width=6.5cm,height=6.0cm]{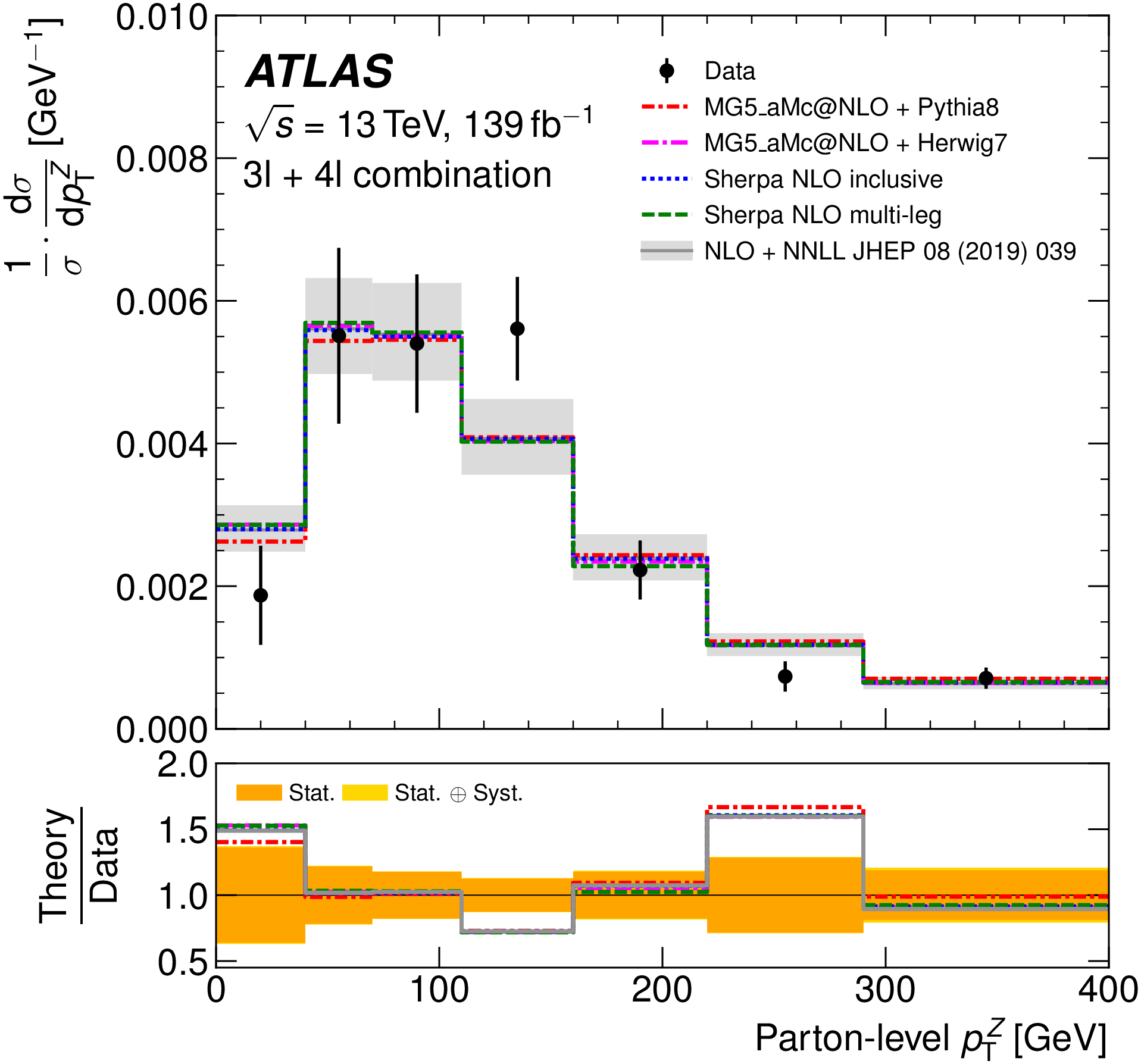}
\caption{Comparison of normalised unfolded particle- and parton-level distribution of the transverse momentum of the Z boson in observed data from ATLAS~\cite{ATLASttZ2021} with Theoretical expectations obtained from different generators: Sherpa 2.2.1~\cite{Sherpa221} generator at NLO QCD accuracy using either multi-leg or inclusive setups and $\mathrm{MadGraph5\_aMC@NLO}$~\cite{MadGraph5} at NLO QCD accuracy interfaced with either the Pythia~\cite{pythia8} or Herwig~\cite{herwig} parton shower models. \label{fig_ATLAS_ttZ_pTZ_norm}}
\end{figure}

The fiducial volumes in which the measurements are made are defined using particle and parton-level objects, respectively, with a selection designed to be as close to the selection used in the inclusive measurement as possible. The background contributions are estimated in the same way as for the inclusive cross-section measurement. The WZ/ZZ plus jets background normalisation is corrected using normalisation factors obtained in a fit of the inclusive cross-section, based on the 3 and 4 lepton regions. All backgrounds are subsequently subtracted from the data. Several observables are measured, with most resulting in agreement between the background subtracted, unfolded data and the NLO simulation with which it is compared. Figure \ref{fig_ATLAS_ttZ_pTZ_norm} shows the agreement between the unfolded particle-level data distribution of the Z-boson transverse momentum and the four theoretical predictions.

\begin{figure}[H]
\centering
\captionsetup{justification=centering}
\includegraphics[width=6.5cm,height=6cm]{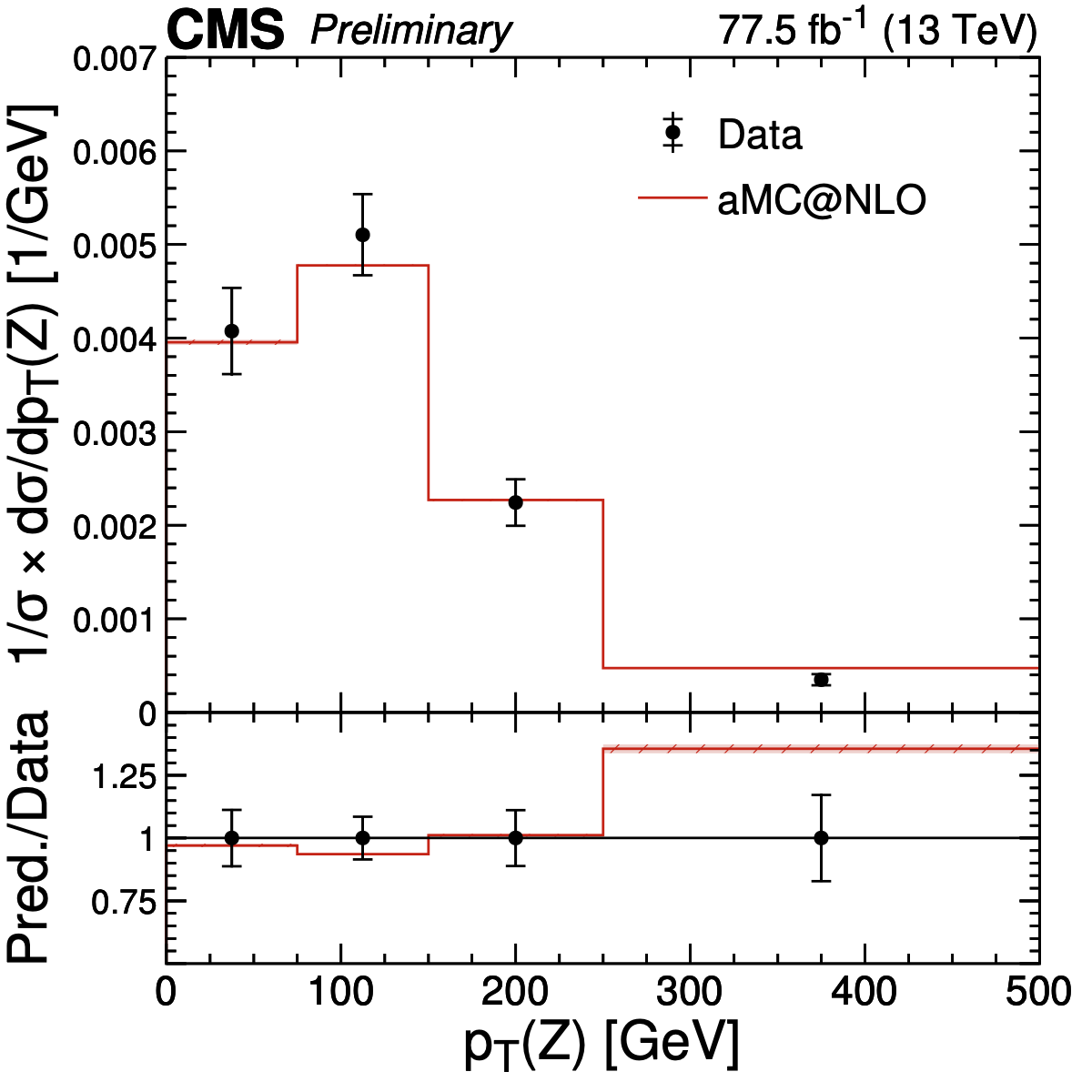}
\caption{Comparison of normalised unfolded parton$-$level distribution of the transverse momentum of the Z boson in observed data from CMS~\cite{CMSttZ2020} with Theoretical expectations obtained from different generators: Sherpa 2.2.1 generator at NLO QCD accuracy using either multi-leg or inclusive setups and $\mathrm{MadGraph5\_aMC@NLO}$ at NLO QCD accuracy interfaced with either the Pythia or Herwig parton shower models. \label{fig_CMS_ttZ_pTZ_parton_norm}}
\end{figure}

The differential cross-section measurement from CMS is performed in the same fiducial volume as defined for the inclusive measurement. Data are unfolded to parton level using the TUnfold package~\cite{tunfold}, which implements a least square fit with a Tikhonov regularisation. The unfolded distribution of the Z-boson transverse momentum is shown in Figure \ref{fig_CMS_ttZ_pTZ_parton_norm} along with the prediction from the $\mathrm{MadGraph5\_aMC@NLO}$ Monte Carlo simulation.

CMS also provide an interpretation of the results in the context of the Standard Model Effective Field Theory (SMEFT) in the Warsaw basis. Anomalous couplings are parametrised by 59 independent Wilson coefficients (WC's) of mass dimension 6, of which 15 are relevant for top-quark interactions. Of these 15, processes involving t-Z interactions can provide competitive constraints on four Wilson coefficients: $c_{tZ}$, $c_{tZ}^{[I]}$, $c_{\Phi t}$ and $c_{\bar{\Phi} Q}$. The first two can induce anomalous EW dipole moments while the second two can induce anomalous neutral-current couplings. The values for these parameters will affect the kinematics and normalisation of processes with such vertices and can therefore be probed using differential distributions of the $\mathrm{\text{t}\bar{\text{t}}Z}$ process. Signal yield predictions for non-zero (and zero = SM point) values of anomalous couplings are simulated in an independent sample at LO accuracy. Ratios of the BSM and SM points in a two-dimensional parton-level plane of the $p_T(Z)$ and $cos\theta^*_Z$ distributions are used to re-weight the nominal SM NLO $\mathrm{\text{t}\bar{\text{t}}Z}$ sample. To validate this procedure, the distributions from the reweighted NLO SM sample and the dedicated LO BSM sample are then compared at various points in the WC parameter space after the full event reconstruction and are found to be in agreement.

\begin{figure}[H]
\centering
\captionsetup{justification=centering}
\includegraphics[width=6.5cm,height=6.0cm]{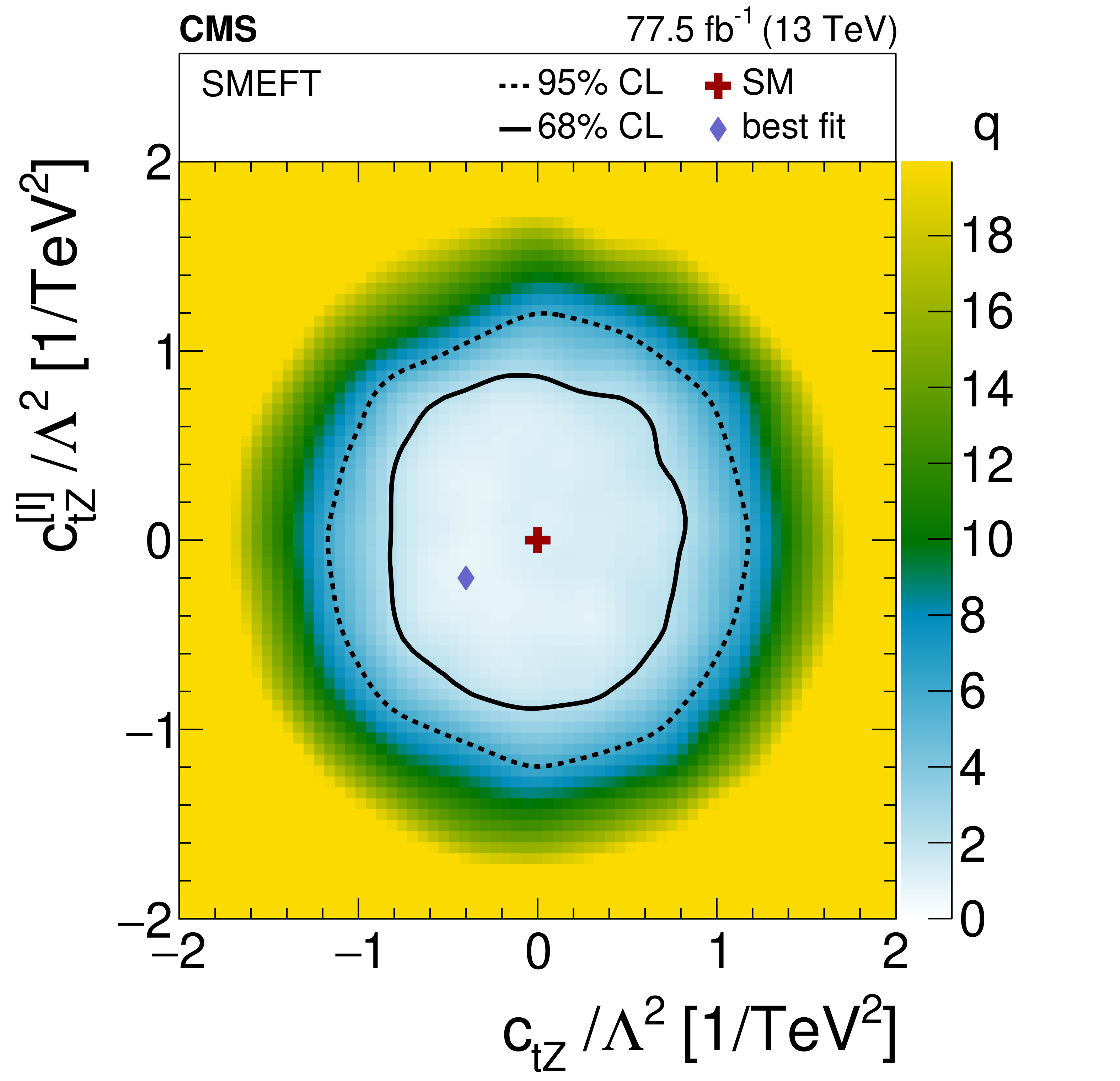}
\includegraphics[width=6.5cm,height=6.0cm]{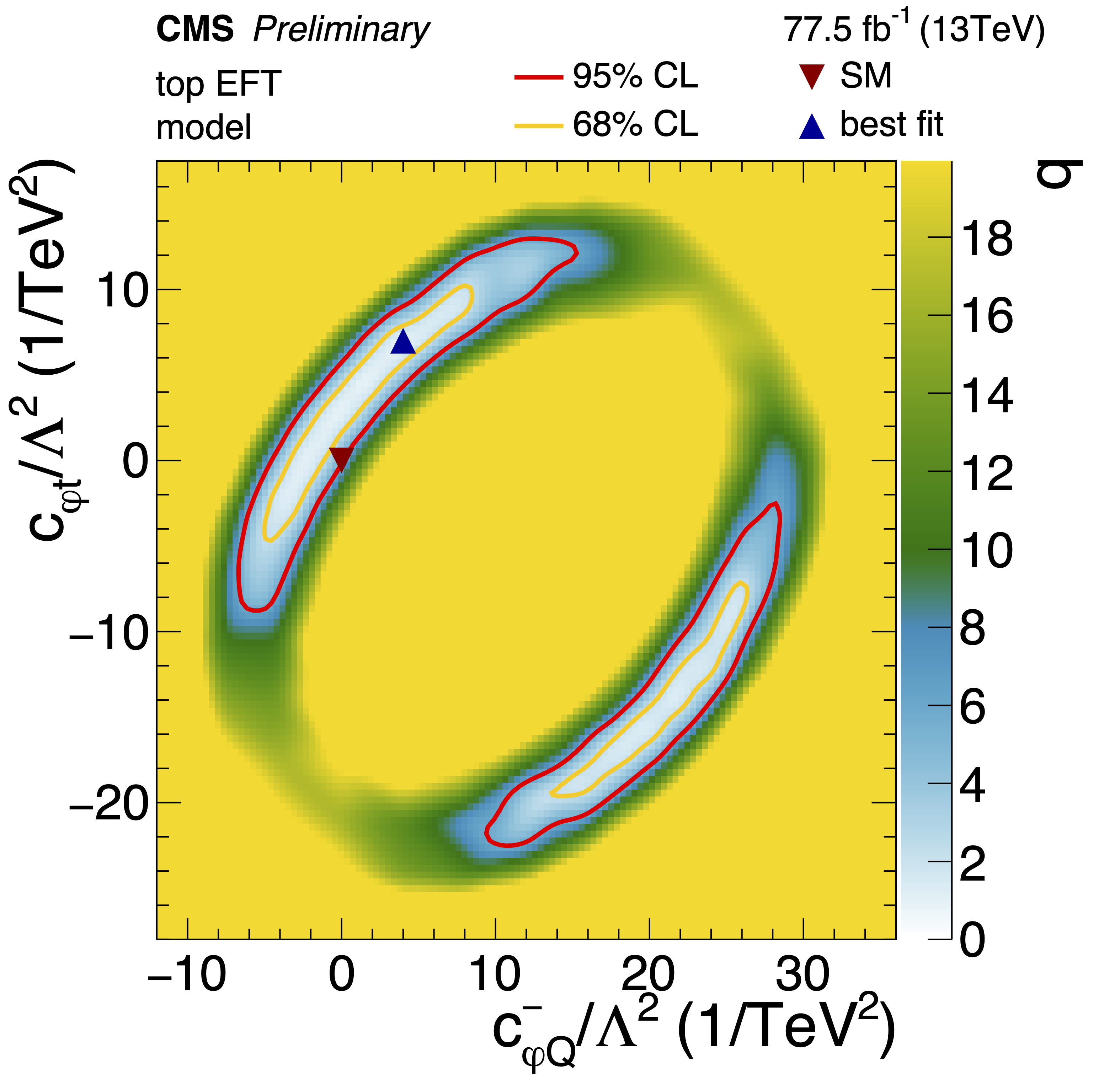}
\caption{Two$-$dimensional scan of $c_{tZ}$ with $c_{tZ}^{[I]}$ and $c_{\Phi t}$ with $c_{\bar{\Phi} Q}$ Wilson coefficients~\cite{CMSttZ2020}. \label{CMS-ttZ_cTZ_cTZI_2Dscan}}
\end{figure}

A binned likelihood function $\mathcal{L}(\theta)$ is constructed from the product of Poisson probabilities and nuisance parameters from the bins in the differential distribution. The values of the nuisance parameters are maximised for each point in the BSM parameter plane to find which point maximises the likelihood. The test statistic

$$
q = -2log(\frac{\mathcal{L}(\hat{\theta})}{\mathcal{L}(\hat{\theta}_{max})})
$$
where $\mathcal{L}(\hat{\theta})$ is the likelihood function which maximises the nuisance parameters at a given BSM point and $\mathcal{L}(\hat{\theta}_{max})$ is the maximised likelihood function at the BSM point with the maximum likelihood. The test statistic $q$ is shown for 1 and 2-dimensional scans of the WCs in Figure \ref{CMS-ttZ_cTZ_cTZI_2Dscan}. For the 1-dimensional scan, all other WCs are fixed to their SM value. All results agree with the SM.

\section{Simultaneous \boldmath{$\mathrm{\text{t}\bar{\text{t}}Z}$} and \boldmath{$\mathrm{t}Zq$} Measurements using Machine Learning Techniques}

To probe the t-Z interaction even further, CMS has constructed a novel analysis~\cite{ttZ_tZq} in which EFT effects on t-Z sensitive processes are targeted using bespoke machine learning algorithms. The analysis targets $\mathrm{\text{t}\bar{\text{t}}Z}$, $\mathrm{tZq}$ and $\mathrm{tWZ}$ processes with at least three leptons and uses multivariate algorithms to exploit the EFT effects in a multi-observable phase-space, creating observables which are optimally sensitive to the effects of EFT operators.

As with the aforementioned measurements, the focus of the measurement is on operators that can affect the couplings between third generation quarks and EW vector bosons. Thus, the same operators are studied but excluding the imaginary component of the complex Wilson coefficient $c_{tZ}^{[I]}$ as it does not conserve CP. Two additional operators are studied however: $c_{tW}$ probing the t-W EW dipole moment and $c_{\Phi Q}^3$ which probes the left-handed SU(2) triplet current operator.

\begin{figure}[H]
\centering
\captionsetup{justification=centering}
\includegraphics[width=13.5cm,height=15.5cm]{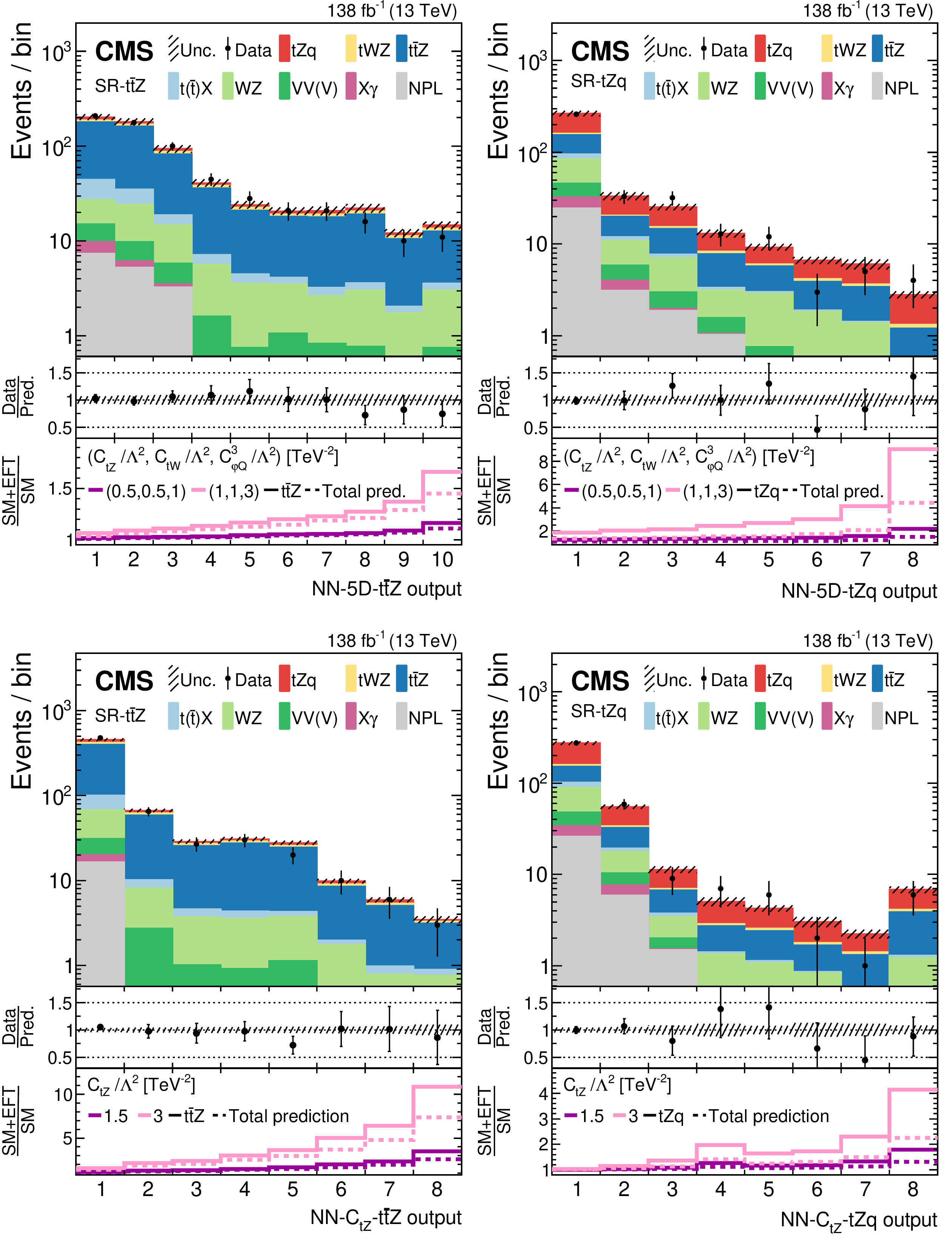}
\caption{Post$-$fit distributions of the EFT neural networks in the $\mathrm{\text{t}\bar{\text{t}}Z}$ and $\mathrm{tZq}$ signal regions from~\cite{ttZ_tZq}. The top row shows the 5D discriminant while the bottom row shows the discriminant trained to target the effects of the $c_{tZ}$ operator. The middle ratio plot demonstrates the data/MC agreement, while the lower ratio demonstrates the increasing impact on the yields in each bin from larger WC~values. \label{CMS-EFTNNs}}
\end{figure}

A multi-classifier is trained to discriminate between the signals and major backgrounds. Separate binary classifiers are trained to discriminate between events generated under the SM and BSM (non-zero WC values) hypotheses. Training datasets are constructed from events randomly sampled from the SM scenario (labelled as background) and BSM scenario (labelled as signal). A novel approach that parameterises the event-weights as a 2{nd} order polynomial is used~\cite{CMSmultileptonEFT}. This makes it possible to smoothly interpolate predictions of the yields in bins of kinematic distributions, between the multitude of different combinations of WC values representing different EFT scenarios. It also allows for the interference between EFT operator amplitudes and either other EFT or the SM amplitudes to be taken into account in the simulation making it possible to exploit these kinematic differences of the various scenarios using a neural network. Separate networks are trained for $\mathrm{\text{t}\bar{\text{t}}Z}$ and $\mathrm{tZq}$ due to their largely different kinematics ($\mathrm{tWZ}$ is not explicitly targeted due to its smaller cross-section and similar kinematics to $\mathrm{\text{t}\bar{\text{t}}Z}$). Training is also performed separately for each operator, along with one training course which targets all five operators simultaneously, allowing for a more global EFT interpretation. Post-fit distributions of the 1D and 5D EFT classifiers are shown in Figure \ref{CMS-EFTNNs}. It is important to note that for larger WC values, the impact on the yield in the more signal-like bins grows stronger, demonstrating how effective these discriminators can be.

\begin{figure}[H]
\centering
\includegraphics[width=12.5cm,height=3.0cm]{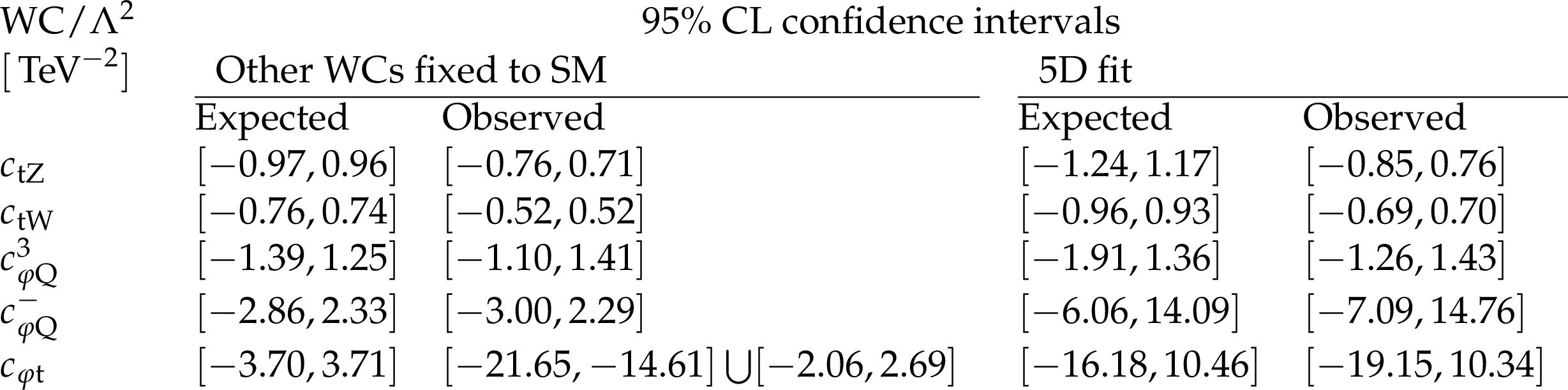}
\caption{95\% CL confidence intervals for the 1D and 5D fits in~\cite{ttZ_tZq}. \label{CMS-EFT-CLs}}
\end{figure}

The distributions of these NN's are fit to data in a maximum likelihood fit, where the likelihood is constructed in the same manner as was described in Section \ref{Section:ttZ}, to establish 68\% and 95\% CL confidence intervals on the values of the WC's. Five 1D scans (one for each operator) of the likelihood are performed, maximising the likelihood in steps of the WC value while fixing the other WCs to zero. Two-dimensional and five-dimensional scans are performed; however, the fit in this case uses the NN trained using distributions sampled from simultaneous variations of the 5 WC. The 95\% CL confidence intervals for the 1D and 5D fits are shown in Figure \ref{CMS-EFT-CLs}.

\begin{figure}[H]
\centering
\captionsetup{justification=centering}
\includegraphics[width=13.8cm,height=6.0cm]{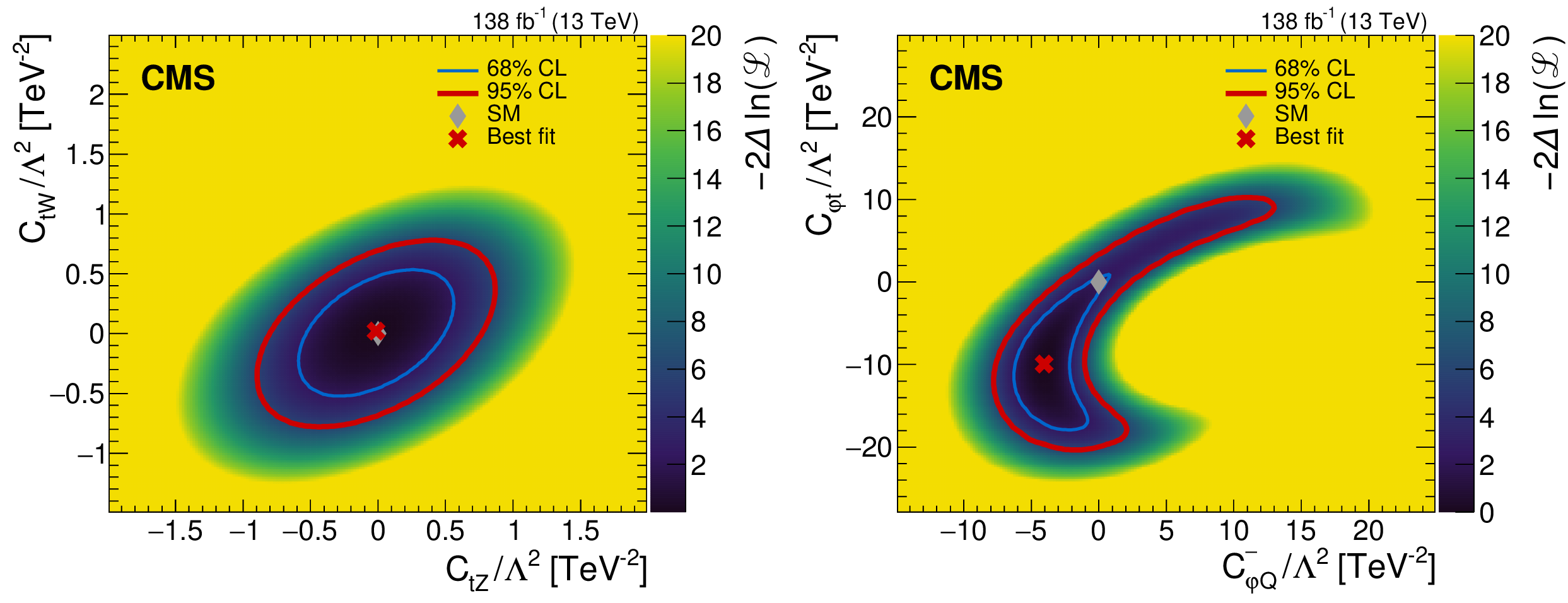}
\caption{95\% CL confidence intervals for the 2D fits in~\cite{ttZ_tZq}. \label{CMS-EFT-2D}}
\end{figure}

Results of the 2D scans to compare with Figure \ref{CMS-ttZ_cTZ_cTZI_2Dscan} are shown in Figure \ref{CMS-EFT-2D}. One can see very competitive results are obtained for common operators. All reported WC values agree with their expected SM value.

\section{\boldmath{$\mathrm{\text{t}\bar{\text{t}}W}$} Measurements}
\begin{figure}[H]
    \centering
    \captionsetup{justification=centering}
    \includegraphics[scale=.5]{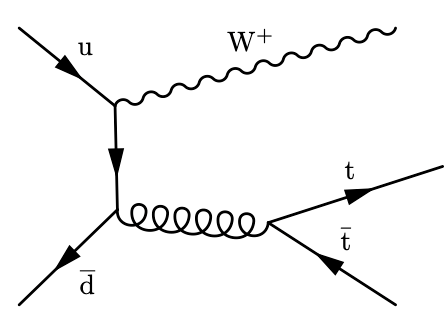}
    \includegraphics[scale=.5]{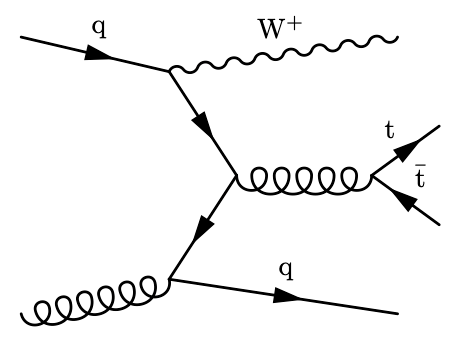}\\
    \includegraphics[scale=.5]{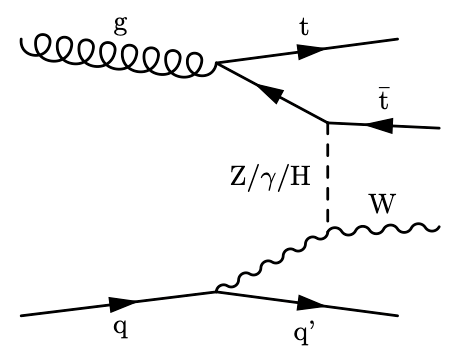}
    \caption{\textls[-15]{Leading-order (top left) and next-to-leading-order (top right and bottom) Feynman diagram for the $\mathrm{\text{t}\bar{\text{t}}W}$ process. The last diagram is an example of the sub-leading electroweak corrections.} \label{ttW-LO}}
\end{figure}

The $\mathrm{\text{t}\bar{\text{t}}W}$ process is unique among processes in which the $\mathrm{\text{t}\bar{\text{t}}}$ system is produced with an associated boson. At leading-order the W boson can only be produced in the initial state, as is shown in Figure \ref{ttW-LO}. The dominant contribution to the total amplitude is form quark-initiated processes. The W boson in fact polarises the incoming quarks and subsequently the top-quark pair leading to an enhancement in the decay product asymmetry at LO, exemplifying the need to take special care of spin correlations in any simulation~\cite{ttW_charge_asym}. Furthermore, the dominance of the quark-initiated production also leads to the $\mathrm{\text{t}\bar{\text{t}}W\pm}$ asymmetry, in which $\mathrm{\text{t}\bar{\text{t}}W+}$ production dominates over $\mathrm{\text{t}\bar{\text{t}}W-}$, and is sensitive to the parton density function (PDF) of the proton.

Fixed order calculations of $\mathrm{\text{t}\bar{\text{t}}W}$ at NLO in QCD ($\alpha_S^3\alpha$) have existed for a long time~\cite{ttW_FO_NLO} and have been matched to parton shower~\cite{ttW_NLO_PSmatched,ttW_NLO_ttHimpact}, with NLO EW corrections ($\alpha_S^2\alpha^2$) coming later~\cite{ttZ_theoryXS_3}.

Persistent tensions between the measurements and predictions of the $\mathrm{\text{t}\bar{\text{t}}W}$ cross-section have driven a lot of recent activity in the theory community. Calculations have become increasingly more sophisticated despite the many difficulties that arise when calculating the higher-order corrections for this process.

$\mathrm{\text{t}\bar{\text{t}}W}$ production with an additional parton (e.g., $\mathrm{\text{t}\bar{\text{t}}Wj}$ and $\mathrm{\text{t}\bar{\text{t}}Wjj}$) generate large augmentations to the total cross-section with large NLO corrections as they introduce gluon-initiated production processes~\cite{inclusive_ttW_anatomy}. To merge the matrix elements of these processes with PS machinery, dedicated studies have been performed, with an improved multi-leg matching scheme~\cite{FXFXttW}. 

Calculations at NLO in QCD that account for the next-to-next-to-leading logarithmic (NNLL)~\cite{ttV_NLO_NNLL} effects are  now available as well as NLO QCD with NNLL effects with NLO EWK corrections~\cite{ttWnoFxFx,ttW_NLO_NNLL_NLOEWK}. Sub-leading EW corrections ($\alpha^3\alpha_S$) to $\mathrm{\text{t}\bar{\text{t}}W}$ have in fact been found to have a larger effect than expected (approximately 10\%)~\cite{ttW_large_NLO_EW,ttW_strong_tW_scattering,ttW_Powheg}, primarily due to contributions from amplitudes represented by top-W-boson scattering diagrams.

Recent work has also included calculations of the full NLO cross-section including fixed order corrections and full LO spin correlations of decay products using POWHEG~\cite{ttW_Powheg}. Some emphasis has also been put on the need for off-shell calculation which culminated in full off-shell calculations at NLO in QCD~\cite{ttW_NLO_offshell_1,ttW_NLO_offshell_2,ttW_NLO_offshell_3}, off-shell calculations incorporating NLO EWK corrections~\cite{ttW_offshell_QCD_EW} and finally the development of procedures to incorporate off-shell effects into NLO+PS procedures~\cite{ttW_offshell_NLO_PS}


As mentioned in Section \ref{Section:ttZ}, $\mathrm{\text{t}\bar{\text{t}}W}$ inclusive cross-section measurements have in the past been simultaneously extracted the $\mathrm{\text{t}\bar{\text{t}}Z}$ cross-section due to the difficulties in disentangling these two rare processes. The previous measurements from CMS used data collected in 2016, selecting events with two or more leptons. Events selected with two leptons of the same sign charge provide the most sensitivity to the $\mathrm{\text{t}\bar{\text{t}}W}$ process. The inclusive cross-section was measured to be $\sigma_{\mathrm{\text{t}\bar{\text{t}}W}} = 0.77 ^{+0.12}_{-0.11}\mathrm{stat.} ^{+0.13}_{-0.12}\mathrm{syst.} $\, pb with an observed (expected) significance of 5.3 (4.5) standard deviations~\cite{CMSttZ2018}. ATLAS made a similar measurement, extracting a cross-section value of $\sigma_{\mathrm{\text{t}\bar{\text{t}}W}} = 0.87 \pm 0.13\mathrm{stat.} \pm 0.14 \mathrm{syst.} $\, pb and an observed (expected) significance of 4.3 (3.4)~\cite{ATLASttZ2019}.

With the full Run 2 dataset available CMS has performed a new analysis that independently measures the inclusive $\mathrm{\text{t}\bar{\text{t}}W}$ cross-section in the two lepton (same-sign) and three or more lepton channels. Although the much larger dataset significantly reduces the statistical uncertainty, new techniques have been developed to reduce the systematic uncertainty from $16\%$ in the 2016 measurement to $6\%$. One of the key developments was a new multivariate analysis (MVA) algorithm designed to distinguish between leptons from the decays of W bosons (prompt leptons) and leptons originating in either the decay of heavy quarks (b or c quarks) or misidentified hadronic jets (non-prompt leptons). Although non-prompt leptons are generally easy to distinguish from prompt leptons, when background processes are large enough, they will still produce many objects with lepton-like signatures, such that further steps are needed to reduce their contribution to a signal region. The non-prompt background in this analysis primarily stems from the $\mathrm{\text{t}\bar{\text{t}}}$ process. The new MVA algorithm brings a large improvement in the signal efficiency of the analysis compared with when a cut-based identification method was used in the previous iteration.

In the same-sign dilepton category, a multi-class deep neural network (DNN) is used to discriminate between signal and background using kinematic distributions of the jets and leptons in the event. The network is trained to distinguish between four processes: $\mathrm{\text{t}\bar{\text{t}}W}$, non-prompt lepton backgrounds (modelled using $\mathrm{\text{t}\bar{\text{t}}}$ simulation), $\mathrm{\text{t}\bar{\text{t}}Z}$ or $\mathrm{\text{t}\bar{\text{t}}H}$, and $\mathrm{\text{t}\bar{\text{t}}}\gamma$. The distribution of the $\mathrm{\text{t}\bar{\text{t}}W}$ output node provides an optimally discriminating variable.

A likelihood function is built from the Poisson probabilities to obtain the observed yields in bins of the discriminating variables in several event categories, with terms incorporating the various uncertainties and the correlations. A binned profile likelihood fit to the observed data is then performed using predicted signal and background distributions simultaneously in all event categories.

In the dilepton channel events are categorised according to the selected leptons' flavour and charge. The DNN $\mathrm{\text{t}\bar{\text{t}}W}$ output node is the discriminating observable that is then used in the fit. In the tri-lepton category, events are categorised according to their number of jets, medium b-tags and the charge of the selected leptons, and the tri-lepton mass m($3\ell$) is used as the fit observable.

\begin{figure}[H]
    \centering
    \captionsetup{justification=centering}
    \includegraphics[width=8cm,height=8cm]{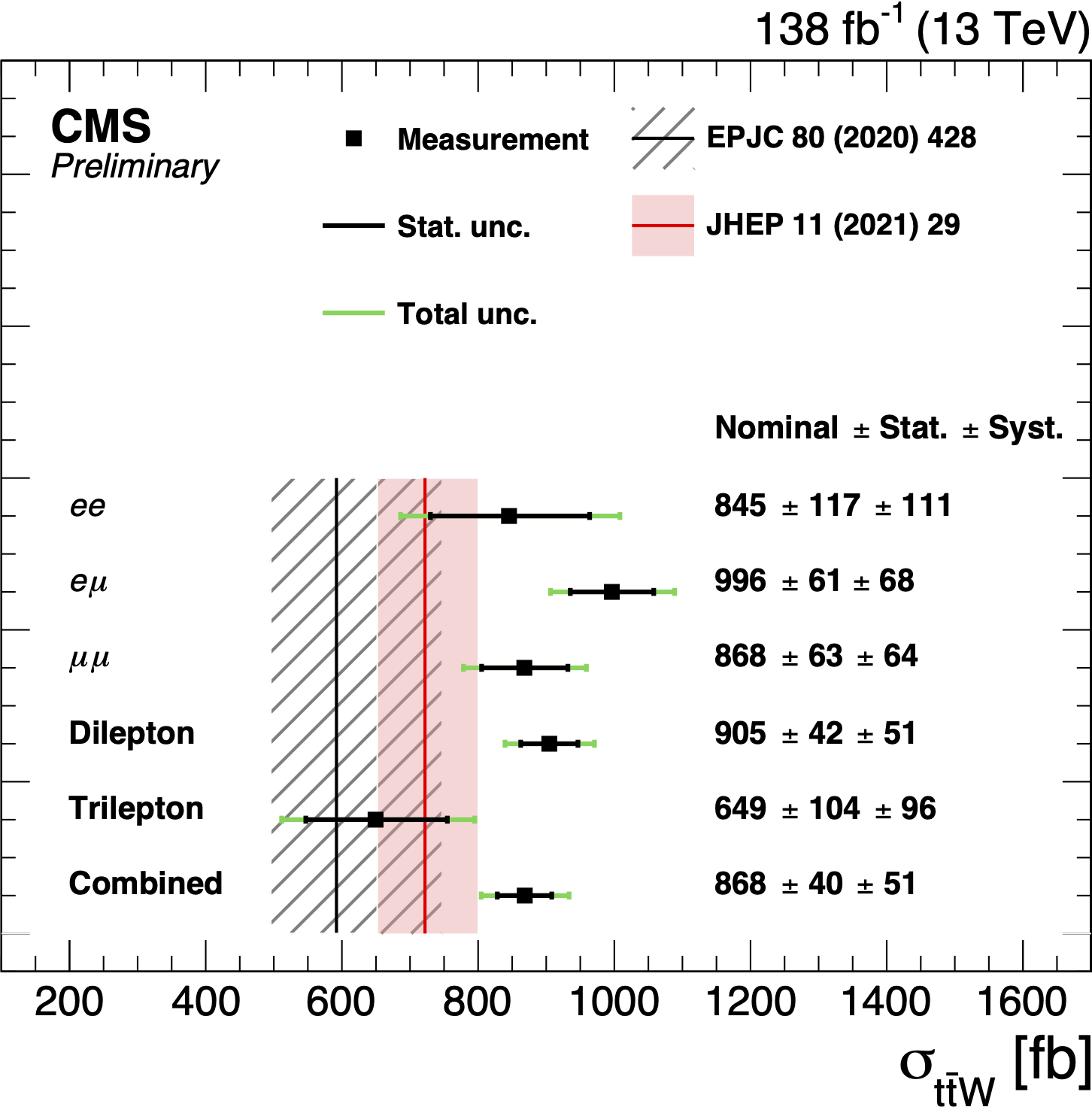}
    \caption{Measurements of the inclusive $\mathrm{\text{t}\bar{\text{t}}W}$ cross$-$section~\cite{CMS_ttW_2022}. The combined result is shown with a breakdown of the measurement obtained in the different dilepton and tri$-$lepton channels, as well as the measurement obtained in the different lepton flavour categories of the dilepton channel. The black inner error bar indicates the statistical uncertainty, while the green outer error bar represents the full systematic plus statistical uncertainty. The measurements are compared with two SM predictions. The prediction shown by the black line is from Ref.~\cite{ttWnoFxFx} while the prediction represented by the red line comes from Ref.~\cite{FXFXttW} and includes FxFx predictions. The central lines of these two vertical lines represent the nominal prediction, while the band represents the combined uncertainty from the scale and PDF theory variations in the calculation. \label{CMS-ttW-incl-XS}}
\end{figure}

 The inclusive $\mathrm{\text{t}\bar{\text{t}}W}$ production cross-section is measured to be $\sigma_{ \mathrm{\text{t}\bar{\text{t}}W} } = 868 \pm 40 \mathrm{(stat)}$  $ ^{+52}_{-50} \mathrm{(syst)}$  fb~\cite{CMS_ttW_2022}, which is the most precise measurement to date. A breakdown of the cross-section measurement in the different channels is found in Figure \ref{CMS-ttW-incl-XS} where it is compared with two theory predictions. The SM prediction at NLO+NNLL accuracy with FxFx jet merging represents the latest theory prediction~\cite{FXFXttW} giving a cross-section of $\sigma_{ \mathrm{\text{t}\bar{\text{t}}W} }^{\mathrm{theo}} = 592 ^{+155}_{-96} \mathrm{(scale)}\, \pm 12 \mathrm{(PDF)}\,$fb. Measured and predicted cross-sections are within two standard deviations of one another. The central value of the measurement in data is approximately 1.5 times larger than the comparative theory prediction.

The dominant systematic uncertainties originate from the uncertainty on the luminosity determination, the background estimation of the electron charge misidentification rate and the b-jet identification. All these uncertainties have significantly reduced with respect to the last iteration.

\begin{figure}[H]
    \centering
    \captionsetup{justification=centering}
    \includegraphics[width=7cm,height=7cm]{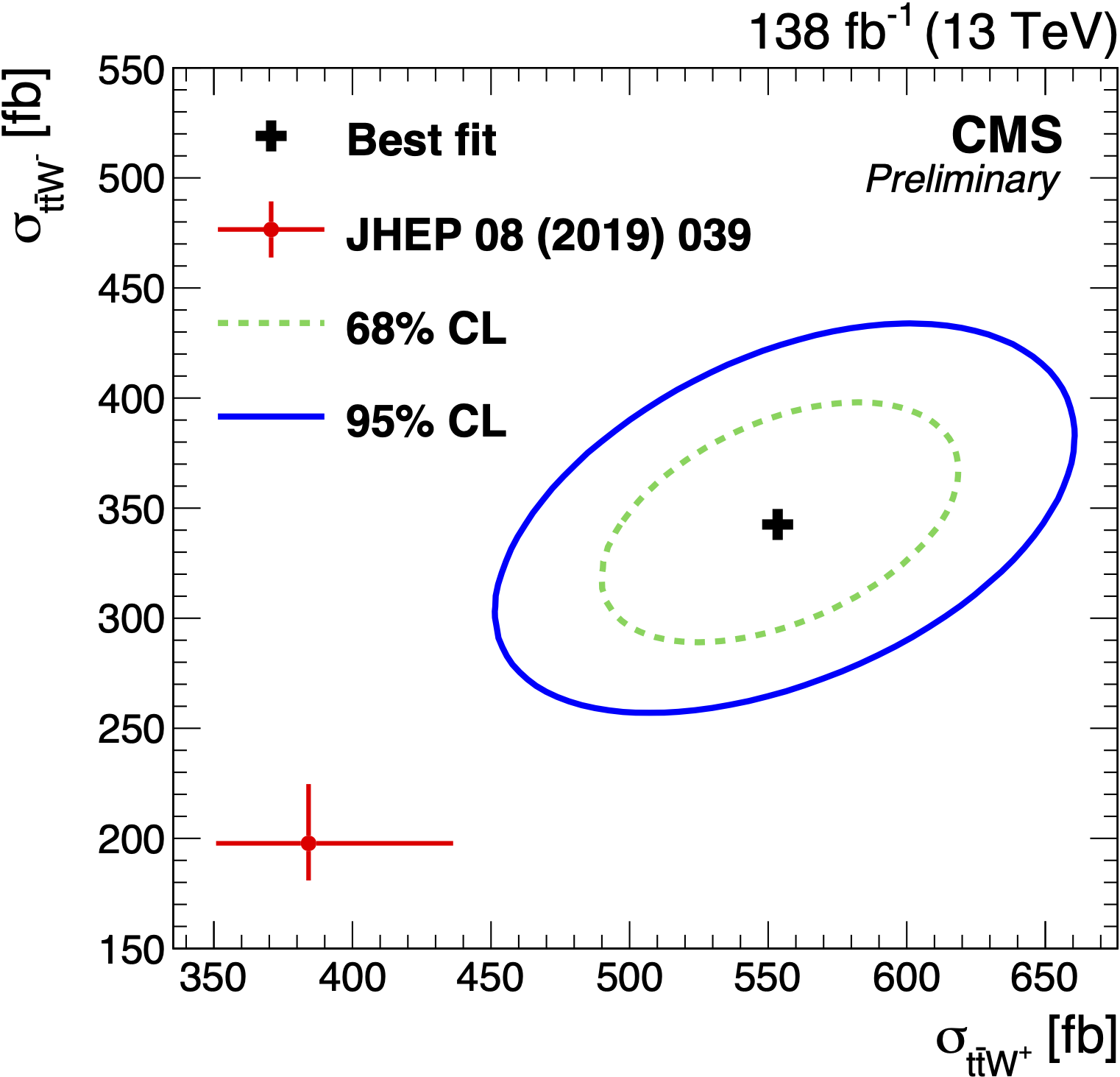}
    \caption{Contours showing the $68\%$ and $95\%$ CL intervals from the likelihood fit in which the $\mathrm{\text{t}\bar{\text{t}}W^+}$ and $\mathrm{\text{t}\bar{\text{t}}W}^-$ processes are measured simultaneously as independent parameters~\cite{CMS_ttW_2022}. The best fit value of the fit is indicated by the black cross, with the theory prediction from Ref.~\cite{ttWnoFxFx} shown by the red cross. The theory prediction included is without the FxFx jet merging. \label{CMS-ttWp-ttWm}}
\end{figure}

\begin{figure}[H]
    \centering
    \captionsetup{justification=centering}
    \includegraphics[width=7cm,height=7cm]{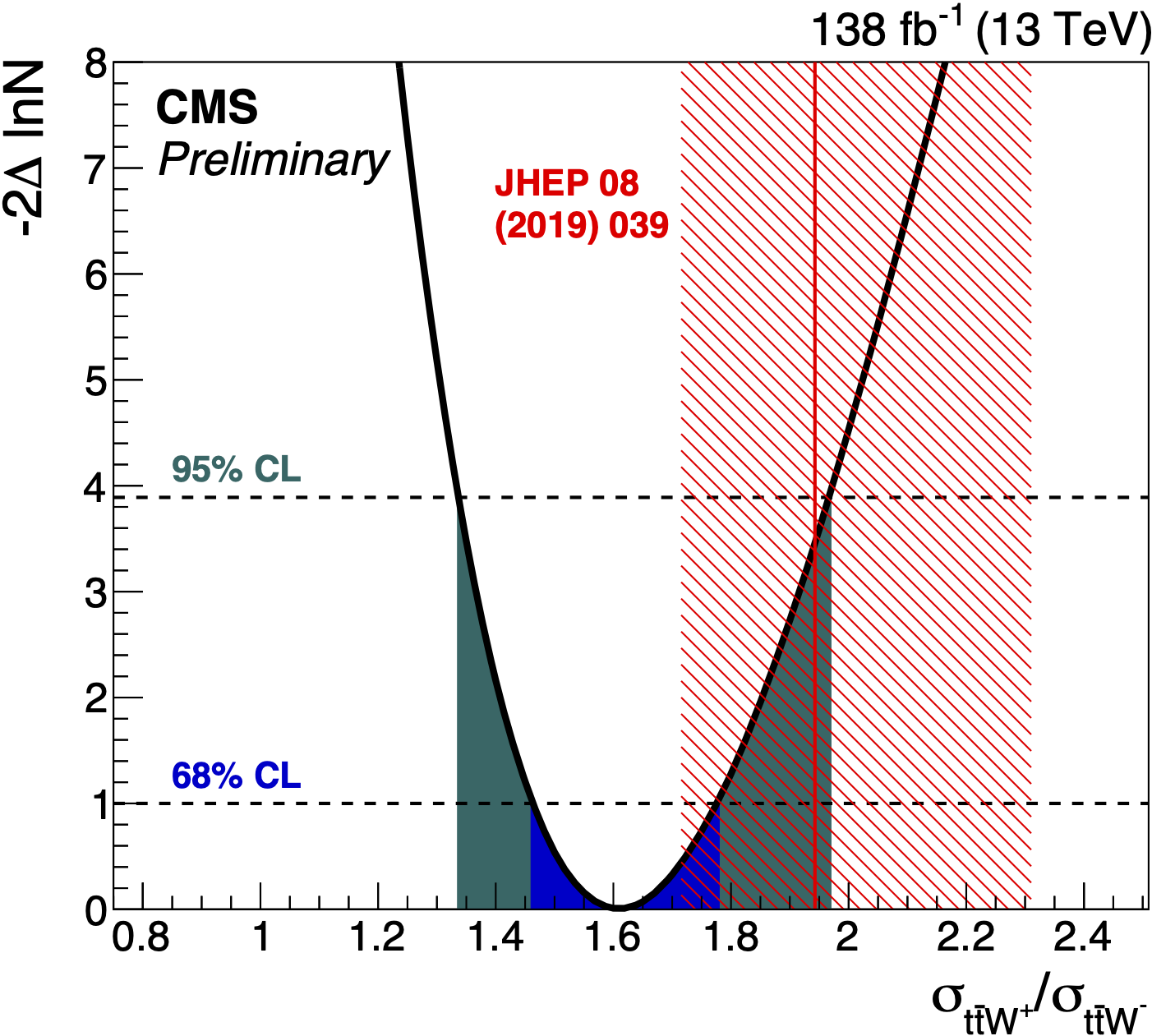}
    \caption{Negative log$-$likelihood scan for values of the ratio of $\mathrm{\text{t}\bar{\text{t}}W^+}$ and $\mathrm{\text{t}\bar{\text{t}}W}^-$ cross-sections. The best fit value is found at the minimum of the curve, while the dashed horizontal lines represent the CL limits~\cite{CMS_ttW_2022}. The red line and hatched band represent the central value and total uncertainty of the theory prediction without the FxFx merging in Ref.~\cite{ttWnoFxFx}. \label{CMS-ttWp-ttWm-ratio}}
\end{figure}

A simultaneous measurement of the $\mathrm{\text{t}\bar{\text{t}}W^+}$ and $\mathrm{\text{t}\bar{\text{t}}W}^-$ cross-sections is performed. The results in Figure \ref{CMS-ttWp-ttWm} show that the measured cross-sections are significantly lower than the theoretical prediction. A measurement of the ratio of these two cross-sections is performed, as there are partial correlations between the systematic uncertainties of the two cross-sections that are reduced when measuring the ratio directly. This measurement is shown in Figure \ref{CMS-ttWp-ttWm-ratio} to also be low in the theoretical prediction, but in agreement within the~uncertainties.

\unskip

\section{\boldmath{$\mathrm{\text{t}\bar{\text{t}}}\gamma$} Measurements}
Studies of the $\mathrm{\text{t}\bar{\text{t}}}\gamma$ production process probe the behaviour of the $\mathrm{t}\gamma$ electroweak coupling. The cross-section is sensitive to new physics that can occur via anomalous dipole moments of the top. Differential measurements provide additional sensitivity to said modifications that may affect spectra more or less in a particular kinematic regime. Such measurements typically compare  state-of-the-art theory predictions with data to stress test the SM, and can be used to probe for BSM physics in a mode independent way.

The $\mathrm{\text{t}\bar{\text{t}}}\gamma$ process is the rarest of the processes discussed in this review. Despite the small production cross-section, the associated production of a photon creates a very distinctive signature that manifests as an isolated energy deposit in the electromagnetic calorimeter without any associated tracks in the silicon tracker. This, along with several jets and leptons, facilitates a high purity event selection. As a result, evidence of this process was first seen by the CDF Collaboration in $\sqrt{s}=1.96$\,TeV collisions~\cite{ttyCDF}. It was subsequently observed at the LHC by the ATLAS Collaboration in $\sqrt{s}=7$\,TeV proton–proton collisions~\cite{ttyATLAS7tev} and has been measured by both ATLAS and CMS in $\sqrt{s}=8$\,TeV~\cite{ttyATLAS8tev,ttyCMS8tev}.

Both collaborations have now also measured this process using $13$TeV pp collisions. The first measurement at this energy scale was performed by the ATLAS collaboration in leptonic final states~\cite{ATLAStty13tev_ljets} using a luminosity of $36.1~\mathrm{fb^{-1}}$, which accounts for a subset of the full Run 2 dataset. Subsequently, measurements using the full Run 2 dataset of $138~\mathrm{fb^{-1}}$ were performed by CMS in the single-lepton~\cite{CMStty13tev_ljets} and dilepton~\cite{CMStty13tev_dilep} final states. Similarly, ATLAS uses a full Run 2 dataset of $139~\mathrm{fb^{-1}}$ use  targeting the dilepton ($\mathrm{e}\mu$)~\cite{ATLAStty13tev_emu} final state~only.

\begin{figure}[H]
    \centering
    \captionsetup{justification=centering}
    \includegraphics[width=6cm,height=4.0cm]{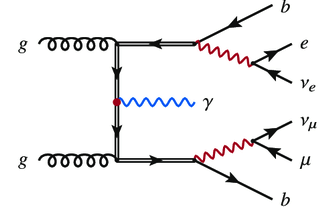}
    \caption{Leading-order Feynman diagram for the $\mathrm{\text{t}\bar{\text{t}}}\gamma$ process. Each diagram demonstrates a different production mechanism for the high energy photon in the process. \label{ttgamma-LO}}
\end{figure}

The targeted signals in all analysis includes the processes demonstrated in Figure \ref{ttgamma-LO}, in which the photon not only originates from the top-quark decay but also the charged fermions radiated from the decay products of the top quark, and from the incoming parton. No attempt to differentiate between the sources is made, but requirements on the photon kinematics are implemented to suppress photons from the top-quark decay products.

\begin{figure}[H]
    \centering
    \captionsetup{justification=centering}
    \includegraphics[width=6.0cm,height=4.0cm]{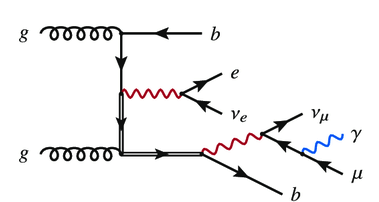}
    \caption{Leading-order Feynman diagram for the $\mathrm{tW}\gamma$ process. Red gauge boson lines represent W bosons while blue gauge boson lines represent photons~\cite{ATLAStty13tev_emu}. \label{tWgamma-LO}}
\end{figure}

ATLAS performed its latest dilepton measurement in the $\mathrm{e}\mu$ channel only, due to the clean final state and small background contribution. This enables an analysis strategy without having to implement complicated MVAs to discriminate signal from background thus simplifying the subsequent comparison with theoretical calculations. In particular, the analysis targets a comparison with the  $\mathrm{pp \rightarrow bWbW}\gamma$ calculation in reference~\cite{hardphoton_ttgamma,tWy_precise}. The calculation includes all resonant and non-resonant diagrams, interference and off-shell effects of the top quarks and W bosons, meaning the signal considered combines both resonant $\mathrm{\text{t}\bar{\text{t}}}\gamma$ and non-resonant $\mathrm{tW}\gamma$ production as demonstrated in Figure \ref{tWgamma-LO}.

Each analysis defines its own signal region at the detector-level where events are selected with exactly one photon, at least one b-tagged jet, and a channel-dependent number leptons (electrons or muons) and jets. After the full event selection, the persisting backgrounds can be broadly categorised as coming from four sources, three of which originate from events in which the photon or the lepton has been misidentified. Each measurement then defines a fiducial volume using particle-level objects, except for the ATLAS dilepton $\text{e}\mu$ measurement, which uses parton-level objects. A summary of the different fiducial volumes is shown in Table \ref{ttgamma_fidsel_tab}.

\begin{table}[H] 
\caption{Table outlining the fiducial selections made in each analysis. All selections are made on particle-level objects except for the ATLAS dilepton measurement. Additional requirements ensure leptons (in all cases, only electrons and muons are considered) are isolated and that the lepton energy incorporates that of radiated photons. Additional photon requirements also ensure isolation and that it does not originate from hadronic decay. Additional vetoes are applied to events in which leptons and photons are produced in close proximity. In particle-level selections, b-jets are defined using ghost-matching~\cite{ghost_association}. Leptons in the parton-level definition are required to come from the W-boson decay. Superscripts $1$ and $2$ refer to objects ordered by transverse momentum from highest to lowest. \label{ttgamma_fidsel_tab}}
\newcolumntype{C}{>{\centering\arraybackslash}X}
\begin{tabularx}{\textwidth}{CCCCCC}
\toprule
\textbf{Experiment }       &  \textbf{Final State}       &   \textbf{Photon} 	&        \textbf{Leptons (e/$\boldsymbol{\mu}$)}  	&   \textbf{Jets}   &   \textbf{b-jets}\\
\midrule
ATLAS~\cite{ATLAStty13tev_ljets} & $N_{\ell}=1(=2)$, $N_{\gamma}=1$, $N_j\geq4$($\geq2$), $N_b \geq 1$ &  $p_T>20$\,GeV, $|\eta|<2.37$	& $p_T>25$\,GeV, $|\eta|<2.5$ & $p_T>25$\,GeV, $|\eta|<2.5$ & $p_T>25$\,GeV, $|\eta|<2.5$\\
\midrule
ATLAS~\cite{ATLAStty13tev_emu} (parton level) & $N_{e}=1$, $N_{\mu}=1$, $N_{\gamma}=1$, $N_b=1$ & $E_T>20$\,GeV, $|\eta|<2.37$	& $p_T>25$\,GeV, $|\eta|<2.5$ & $p_T>25$\,GeV, $|\eta|<2.5$  & $p_T>25$\,GeV, $|\eta|<2.5$ \\
\midrule
CMS~\cite{CMStty13tev_ljets} &$N_{\ell}=1$, $N_{\gamma}=1$, $N_j\geq3$, $N_b \geq 1$ & $p_T>20$\,GeV, $|\eta|<1.44$	& $p_T^e>35$\,GeV, $p_T^{\mu}>35$\,GeV, $|\eta|<2.37$	& $p_T>30$\,GeV, $|\eta|<2.4$  & $p_T>30$\,GeV, $|\eta|<2.4$\\
\midrule
CMS~\cite{CMStty13tev_dilep} & $N_{\ell}=2$(OS), $N_{\gamma}=1$, $N_b\geq1$ & $p_T>20$\,GeV, $|\eta|<1.44$ & $p_T^{1}>25$\,GeV, $p_T^{2}>15$\,GeV, $|\eta|<2.4$ & $p_T>30$\,GeV, $|\eta|<2.4$ &  $p_T>30$\,GeV, $|\eta|<2.4$\\
\bottomrule
\end{tabularx}
\end{table}

Events in which the selected photon candidate originates from a misidentified jet or non-prompt photon from the decay of a hadron make up the hadronic-fake background. The main process contributing to this background is $\mathrm{\text{t}\bar{\text{t}}}$ where one of the jets in the final state is misidentified as a photon. All analyses use data-driven methods to derive scale factors in regions enriched with the hadronic-fake background which are then applied to the simulated hadronic-fake background prediction in the signal region.

Events in which the selected photon candidate originates from an electron make up the electron-fake background. This is the dominant background source in the dilepton channels. Electron-to-photon fake rates are measured using the tag-and-probe method in control regions using the $\mathrm{Z\rightarrow e e}$ process. The fake rate scale factors are determined by taking the ratio between the fake rate measured in the data and simulation in bins of ${p_T}$ and $\eta$.

Additionally, the backgrounds in which one or more leptons result from either a jet or a non-prompt lepton from heavy-flavour decays (fake-lepton) are estimated directly from data, contributing mainly to the single-lepton channel. The main contribution to this background comes from SM processes in which jets are produced uniquely through the strong interaction i.e., QCD events. The photon in such events can be either prompt or fake. The background contributions from events with a prompt photon, excluding signal events and fake-lepton backgrounds with a prompt photon, are estimated using simulated samples. These include W$\gamma$, Z$\gamma$, single-top+$\gamma$, diboson, and $\mathrm{\text{t}\bar{\text{t}}V}$.

All the analyses discussed report inclusive and differential cross-sections measured in fiducial volumes defined according to the kinematics of the final state particles. Differential distributions of certain variables provide information on specific aspects of the $\mathrm{\text{t}\bar{\text{t}}}\gamma$ process. Photon kinematics such as its $p_T$ and $\eta$ are sensitive to the coupling between the top quark and photon. Distributions of the angular separation between the photon and the top quarks decay products are sensitive to the origin of the photon. Furthermore, studying observables that do not involve the photon provide information on the $\mathrm{\text{t}\bar{\text{t}}}$ system itself.

\subsection{Inclusive Cross-Section Measurements}
The latest ATLAS measurement of the inclusive cross-section in the single-lepton channel also includes a simultaneous measurement on the dilepton channel~\cite{ATLAStty13tev_ljets}. This measurement was performed using a smaller dataset collected in 2016 only consisting of $36.1~\mathrm{fb^{-1}}$, somewhat smaller than the more recent ATLAS dilepton $\text{e}\mu$ measurement~\cite{ATLAStty13tev_emu} that will be described later. Using a neural network to discriminate the $\mathrm{\text{t}\bar{\text{t}}}\gamma$ signal from backgrounds at detector-level, this distribution was then used as the input distribution to a profiled likelihood fit in which the fiducial cross-section is extracted. Several fits are performed, either independently fitting to the data in each channel or fitting to the data in each channel simultaneously. A correction factor for the signal efficiency and event migration into the fiducial region is also used when quoting the results. The measured inclusive fiducial cross-section measurements from~\cite{ATLAStty13tev_ljets} are found to be
\begin{align*}
    &\sigma_{fid}^{SL} = 521 \pm 9 \mathrm{(stat)} \pm 41 \mathrm{(sys)\,fb}\\
    &\sigma_{fid}^{DL} = 69 \pm 3 \mathrm{(stat)} \pm 4 \mathrm{(sys)\,fb}
\end{align*}

\begin{figure}[H]
    \centering
    \captionsetup{justification=centering}
    \includegraphics[width=9.5cm,height=8cm]{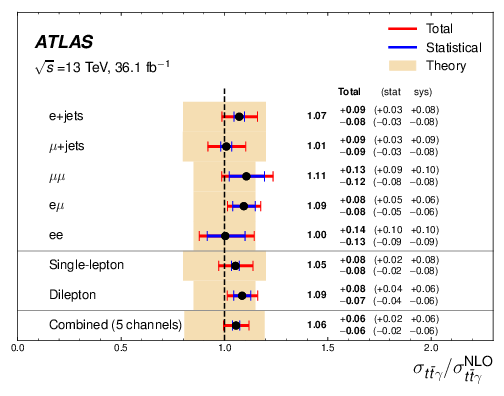}
    \caption{Inclusive $\text{t}\bar{\text{t}}\gamma$ production cross-section measurements by ATLAS in leptonic channels~\cite{ATLAStty13tev_ljets}. The NLO prediction from theory is shown in the dashed vertical line, with the uncertainty shown in the beige band. The measured values in data are represented by the black points, where the associated total and statistical uncertainties are shown in the red and blue lines, respectively. Results in each of the different lepton flavour channels are also shown. \label{ATLAS-ttgamma-leptonic-incl-XS}}
\end{figure}

A breakdown of the results, normalised to their corresponding NLO SM predictions, can be seen in Figure \ref{ATLAS-ttgamma-leptonic-incl-XS}. In the single-lepton channel, the dominant uncertainties are related to the estimates of the jet energy and resolution scales as well as the background modelling, which is dominated by $\text{t}\bar{\text{t}}$ modelling, used to model the hadronic and electron-fake backgrounds. In the dilepton channel the uncertainty is still dominated by the statistical uncertainty of the data, with the largest systematic uncertainty coming from the signal and background modelling, which is dominated by $Z\gamma$ modelling.

CMS performs a similar measurement of the inclusive cross-section in the single-lepton channel. The fiducial phase-space is defined at particle level and can be found in Table~\ref{ttgamma_fidsel_tab}. It is the same for both the inclusive and differential measurements. Signal regions are defined at detector-level and are designed to be as close as possible to the fiducial volume as possible. Additionally, orthogonal control regions are defined, enriched in the major backgrounds, are used in a fit to data to constrain associated uncertainties. The observed and expected yields in the signal and control regions along with the systematic uncertainties, are used to construct a binned likelihood function. The likelihood fit performed to extract the inclusive fiducial cross-section is performed separately to the one for the differential measurement. For the inclusive measurement, events in the signal and control regions are first categorised according to the flavour of the lepton. In the control regions, events are further categorised according to the photon transverse momentum, whereas in the signal regions the $M_3$ variable is used. This $M_3$ variable represents the invariant mass of the three jets that maximises their vector $p_T$ sum. Nuisance parameters are assigned to account for the normalisation of the misidentified electron, Z$\gamma$ and W$\gamma$ backgrounds. The resulting fiducial inclusive cross-section measurement~\cite{CMStty13tev_ljets} is found to be
$$
\sigma(\text{t}\bar{\text{t}}\gamma) = 798 \pm 7 \mathrm{(stat)} \pm 48 \mathrm{(syst)\,fb}
$$

\begin{figure}[H]
    \centering
    \captionsetup{justification=centering}
    \includegraphics[width=9.5cm,height=8.0cm]{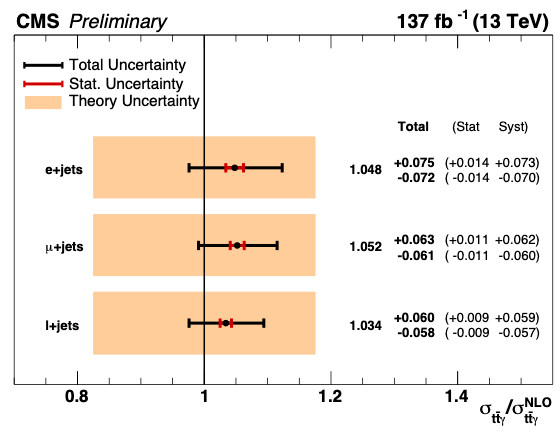}
    \caption{Inclusive $\text{t}\bar{\text{t}}\gamma$ production cross$-$section measurements by CMS in the single$-$lepton channel~\cite{CMStty13tev_ljets}. Results are also shown for the individual lepton flavour channels. \label{CMS-ttgamma-incl-XS}}
\end{figure}

A breakdown of the inclusive measurement in the different channels can be seen in Figure \ref{CMS-ttgamma-incl-XS}. The leading systematic uncertainties according to their post-fit impact on the measured cross-section come from the normalisation of the W$\gamma$ background, the non-prompt background estimation and the integrated luminosity estimation.
\vspace{-6pt}

To extract the inclusive $\mathrm{\text{t}\bar{\text{t}}}\gamma$ fiducial cross-sections in the dilepton channel, the CMS measurement uses a very similar strategy to the single-lepton case, making the two measurements easier to combine. The fiducial phase-space is defined at particle level, for which the full definition can be found in Table \ref{ttgamma_fidsel_tab}. A profile likelihood fit to the photon $p_T$ distribution in data across the three data taking periods of Run 2 is performed. The resulting inclusive fiducial cross-section is found to be~\cite{CMStty13tev_dilep}
$$
\sigma_{fid} = 175.2 \pm 2.5 \mathrm{(stat)} \pm 6.3 \mathrm{(syst)\,fb}
$$

This agrees with the predicted inclusive cross-section of
$$
\sigma_{SM} = 155 \pm 27 \,fb
$$

\begin{figure}[H]
    \centering
    \captionsetup{justification=centering}
    \includegraphics[width=9.5cm,height=7.5cm]{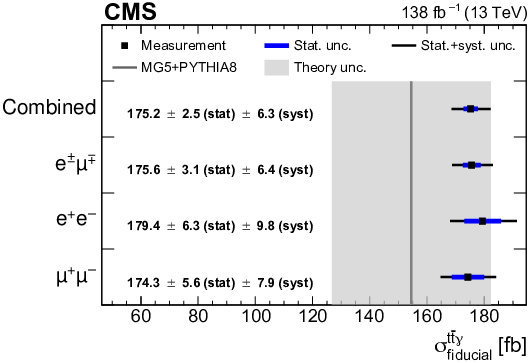}
    \caption{Inclusive $\text{t}\bar{\text{t}}\gamma$ production cross-section measurements by CMS in the dilepton channel~\cite{CMStty13tev_dilep}. Results are also shown for both the combined measurement and the breakdown for the individual dilepton channels. \label{CMS-tty-dilepton-incl-xs}}
\end{figure}

The predicted inclusive cross-section is about 12\% (0.7 standard deviations) lower than the measurement. This is shown in Figure \ref{CMS-tty-dilepton-incl-xs} along with the breakdown of the fit in the individual channels. However, the large theory uncertainties that impact the prediction from Madgraph make it difficult to draw strong conclusions on the agreement between the prediction and the unfolded data. The predicted cross-section is scaled to the NLO $2 \rightarrow 3 \, \text{pp} \rightarrow \text{t}\bar{\text{t}} \gamma$ process, but does not include processes in which the photon is radiated from the final state decay products of the top quark. This is one potential cause of the discrepancy between the results. 

ATLAS measures dilepton fiducial cross-sections in the $\text{e}\mu$ final state using a profile likelihood fit to the $S_T$ distribution (scalar sum of all transverse momenta in the event) in data. This variable provides good separation between the signal and backgrounds. The fiducial volume is defined in Table \ref{ttgamma_fidsel_tab} and is the same for both the inclusive and differential cross-section measurements. The selection mimics that of the theory calculation with which the experimental results are compared~\cite{hardphoton_ttgamma,tWy_precise}. The inclusive cross-section is measured to~be
$$
\sigma_{{fid}} = 39.6 \pm 0.8\, \mathrm{(stat)} ^{+2.6}_{-2.2}\, \mathrm{(syst)}\, \mathrm{fb}
$$

Ref. \cite{ATLAStty13tev_emu}, which agrees with the dedicated theoretical calculation which predicts a value of
$$
\sigma_{{fid}} = 38.50 ^{+0.56}_{-2.18}\, \mathrm{(scale)} ^{+1.04}_{-1.18}\, \mathrm{(PDF)} \, \mathrm{fb}
$$

Refs. \cite{hardphoton_ttgamma,tWy_precise}. As is shown in Table \ref{tty-incl-xs-comparison}, the cross-section measurements all tend to agree with the predicted values at NLO within uncertainties when taking the branching ratios into consideration. Differences in the fiducial cross-sections between the experiments stem from the differences in the fiducial volumes outlined in Table \ref{ttgamma_fidsel_tab}. In particular, the CMS single-lepton fiducial cross-section is measured to be much higher than in ATLAS due to the inclusion of events with three jets and a looser dR() selection.

\begin{table}[H]
\captionsetup{justification=centering}
\caption{Table of the inclusive $\text{t}\bar{\text{t}}\gamma$ production fiducial cross-section measurements from ATLAS and~CMS. 
\label{tty-incl-xs-comparison}}
\newcolumntype{C}{>{\centering\arraybackslash}X}
\begin{tabularx}{\textwidth}{CCC}
\toprule
\textbf{Experiment}	& \boldmath{$\mathrm{\text{t}\bar{\text{t}}}$ \textbf{Decay Channel}} &  \boldmath{$\sigma_{fid}^{\text{t}\bar{\text{t}}\gamma}$} \textbf{(fb)}\\
\midrule
CMS~\cite{CMStty13tev_ljets}		& Single lepton	& 798\\
ATLAS~\cite{ATLAStty13tev_ljets}	& Single lepton	& 521\\
CMS~\cite{CMStty13tev_dilep}		& Dilepton		& 175\\
ATLAS~\cite{ATLAStty13tev_emu}	& Dilepton (e$\mu$) & 39.6\\
\bottomrule
\end{tabularx}
\end{table}

\subsection{Differential Cross-Section Measurements}
\begin{figure}[H]
    \centering
    \captionsetup{justification=centering}
    \includegraphics[width=6.0cm,height=7.0cm]{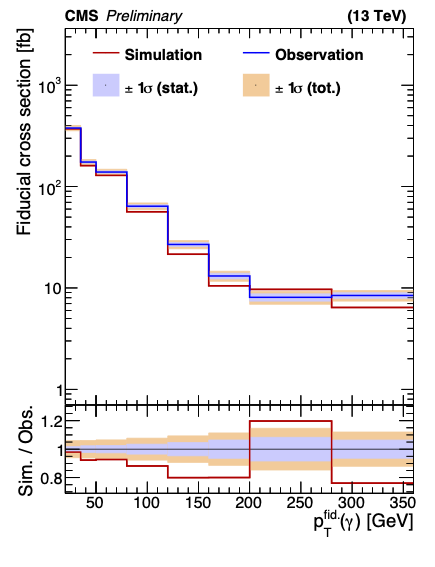}
    \caption{Differential $\text{t}\bar{\text{t}}\gamma$ production cross-section measurements by CMS in the single-lepton channel~\cite{CMStty13tev_ljets}. Results are also shown as a function of the transverse momentum of the photon at particle level. \label{CMS-ttgamma-diff-XS}}
\end{figure}

CMS has reported differential $\mathrm{\text{t}\bar{\text{t}}}\gamma$ fiducial cross-sections in both the single lepton~\cite{CMStty13tev_ljets} dilepton~\cite{CMStty13tev_dilep} channels. The single-lepton publication reports differential fiducial cross-section measurements as a function of the photons $p_T$, $|\eta|$ and the difference in angle between the lepton and photon ($\Delta R(\ell,\gamma)$). Results were obtained simultaneously for the 3 and 4 jet regions, the lepton flavour channels, and the different data taking periods. The same control regions are used as in the inclusive measurement. After the profile likelihood fit, backgrounds are subtracted from the observable distribution in data and subsequently unfolded to particle level. The unfolded differential cross-section is defined in the same fiducial phase-space as the inclusive cross-section. Distributions of the unfolded observables are shown in Figure \ref{CMS-ttgamma-diff-XS} where a comparison with simulations obtained using $\mathrm{MadGraph5\_aMC@NLO}$ interfaced with three different parton shower algorithms is shown. In the bulk of the distribution, the dominant uncertainties are similar to those in the inclusive cross-section measurement. For $p_T(\gamma) > 120$ GeV, the uncertainties in the jet energy scale, photon identification efficiency and colour re-connection modelling are the largest sources of uncertainty.

\begin{figure}[H]
    \centering
    \captionsetup{justification=centering}
    \includegraphics[width=7cm,height=7.0cm]{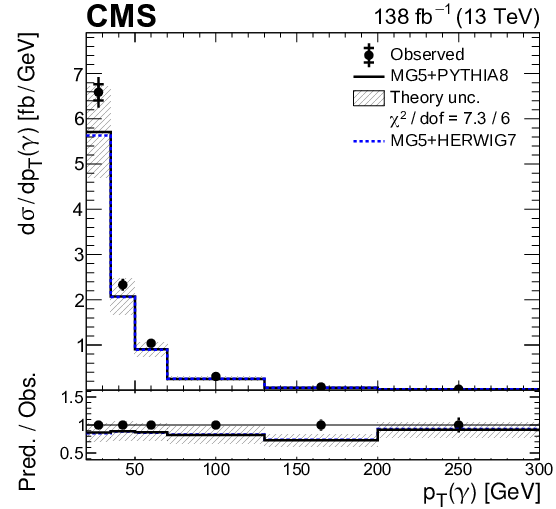}
    \caption{Distribution of the absolute production cross-section of $\text{t}\bar{\text{t}}\gamma$ in the dilepton channel as a function of the $p_T$ of the photon, as measured by the CMS experiment~\cite{CMStty13tev_dilep}. Observed data unfolded to particle level is compared with the predicted distribution from the Madgraph generator with two different parton shower models. Theoretical uncertainties evaluated using the Pythia 8 prediction are shown in the shaded grey bands. \label{CMS-tty-pty-absolute}}
\end{figure}

In the dilepton channel, differential cross-sections are reported with respect to 12~observables that are unfolded to particle level in the same fiducial volume as the inclusive cross-section measurement. These are compared with two predictions using $\mathrm{MadGraph5\_aMC@NLO}$ event generator interfaced with two parton shower simulations: Pythia8 with the CP5 tune~\cite{pythia8_CP5_tune} and Herwig~\cite{MG_HW7} v7.14 with the CH3 tune~\cite{MG_HW_CMStune}. An example of the unfolded distribution of the transverse momentum of the photon at particle level in the dilepton channel is shown in Figure \ref{CMS-tty-pty-absolute}. No significant deviation between the measured distribution and either of the predictions is observed, but due to the size of the theory uncertainties it is once again difficult to come to a conclusion regarding their agreement.

\begin{figure}[H]
    \centering
    \captionsetup{justification=centering}
    \includegraphics[width=7.0cm,height=7.0cm]{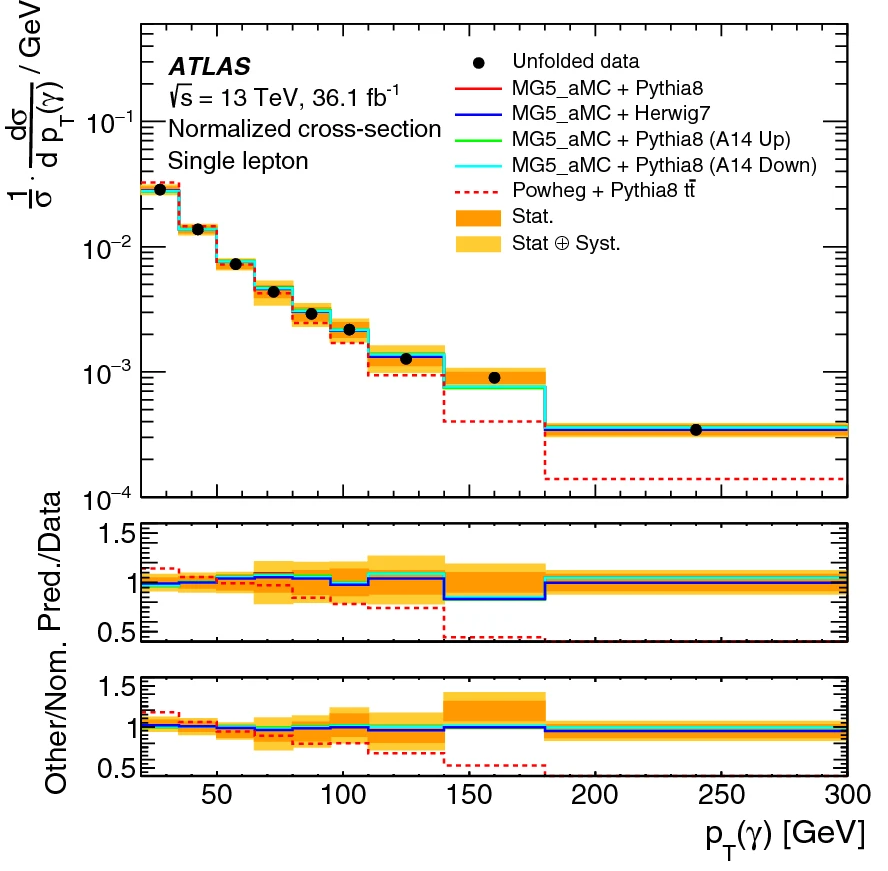}
    \caption{Normalised differential cross-section as a function of the photon transverse momentum~\cite{ATLAStty13tev_ljets}. Unfolded distributions are compared with predictions using the $\mathrm{MadGraph5\_aMC@NLO+Pythia8}$ together with up and down variations of the $\mathrm{Pythia8 A14}$ tune parameters, the $\mathrm{MadGraph5\_aMC@NLO++Herwig 7}$ and $\mathrm{POWHEG+Pythia8}$ $\text{t}\bar{\text{t}}$ where the photon radiation is modelled in the parton shower. \label{ttgamma-leptonic-diff-XS}}
\end{figure}

ATLAS has reported differential cross-section measurements in both the leptonic~\cite{ATLAStty13tev_ljets} and dilepton ($\text{e}\mu$)~\cite{ATLAStty13tev_emu} channels. To extract the distributions, no fit to data is performed. The major backgrounds are subtracted from the data using the estimates outlined earlier after which detector effects are removed using an unfolding procedure which is applied to the observed detector level distributions to obtain the true distribution of the signal at particle or parton level. The differential cross-section is normalised to unity resulting in distributions shown in Figure \ref{ttgamma-leptonic-diff-XS} for~\cite{ATLAStty13tev_ljets}. Absolute differential distributions are also provided and can be found in the paper.

\begin{figure}[H]
    \centering
    \captionsetup{justification=centering}
    \includegraphics[width=7.0cm,height=7.0cm]{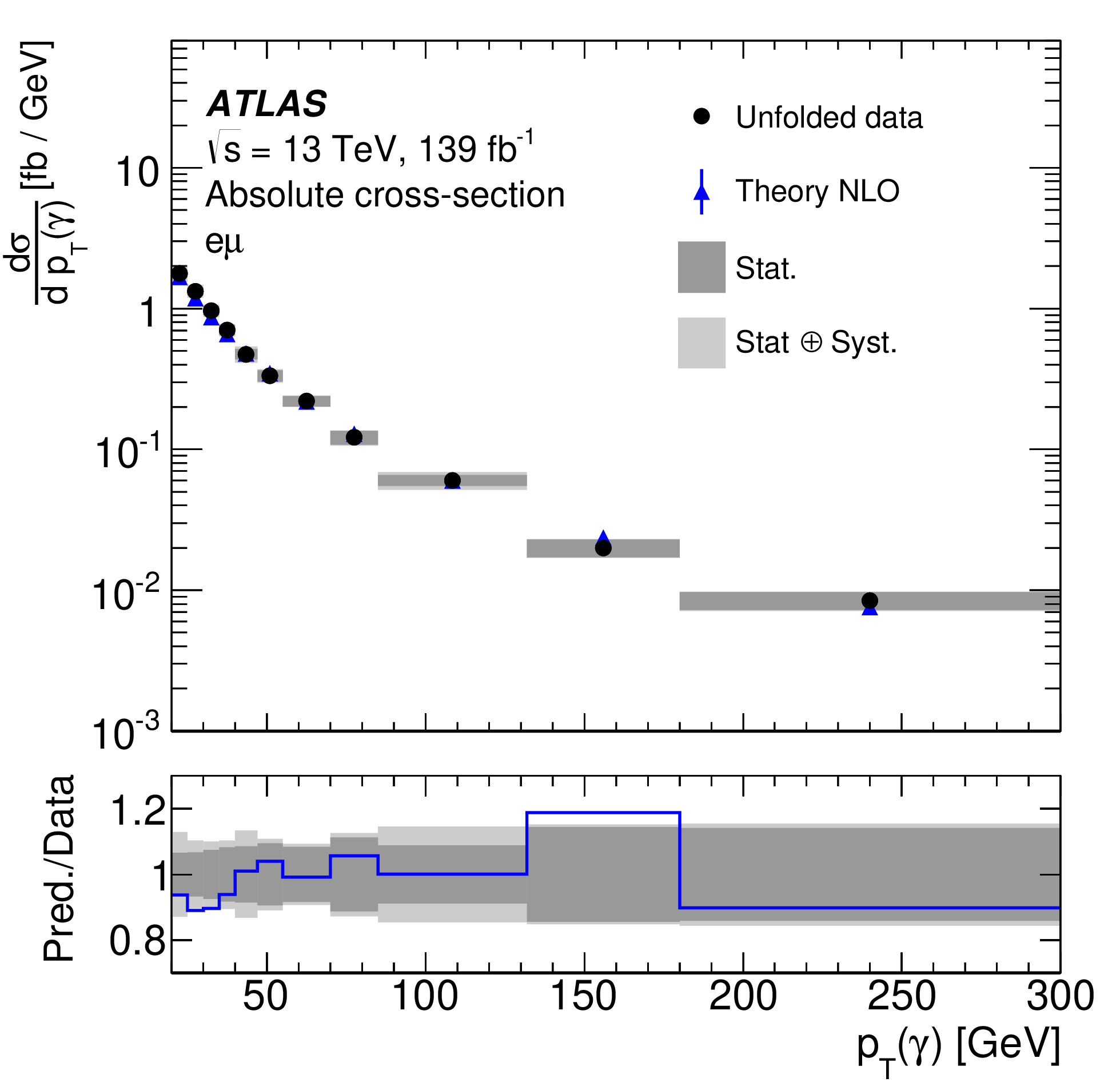}
    \caption{Distribution of the absolute production cross$-$section of $\text{t}\bar{\text{t}}\gamma$ in the $\text{e}\mu$ channel as a function of the $p_T$ of the photon, as measured by the ATLAS experiment~\cite{ATLAStty13tev_emu}. Observed data unfolded to parton level is compared with the predicted distribution from the theoretical prediction from~\cite{hardphoton_ttgamma,tWy_precise}. The systematic and statistical uncertainties are shown in the grey bands. \label{ATLAS-tty-dilepton-pty}}
\end{figure}

In the case of the dilepton ($\text{e}\mu$) channel, ATLAS measures differential cross-sections as a function of a similar set of variables described in the CMS dilepton measurement. Distributions are unfolded to parton level and can therefore be directly compared with the aforementioned theory prediction via both normalised and absolute differential cross-sections. Additionally, a comparison is made with two leading-order simulations using Madgraph interfaced with Pythia or Herwig. A comparison of the parton-level cross-section as a function of the photon $p_T$ in simulation and the unfolded data are shown in Figure \ref{ATLAS-tty-dilepton-pty}. In general, all distributions agree well; however, one trend that was recognised was that the NLO prediction tends to describe most distributions better than the LO prediction.

\vspace{-6pt}

\subsection{EFT Interpretations}
The CMS measurements also provide limits on Wilson coefficients that induce electroweak dipole moments
\vspace{12pt}
\begin{align*}
    &c_{tZ} = Re(-sin\theta_{W}C_{uB}^{33} + cos\theta_{W}C_{uW}^{33})\\
    &c_{tZ}^I = Im(-sin\theta_{W}C_{uB}^{33} + cos\theta_{W}C_{uW}^{33})
\end{align*}

\begin{figure}[H]
    \centering
    \captionsetup{justification=centering}
	\includegraphics[width=6cm,height=6cm]{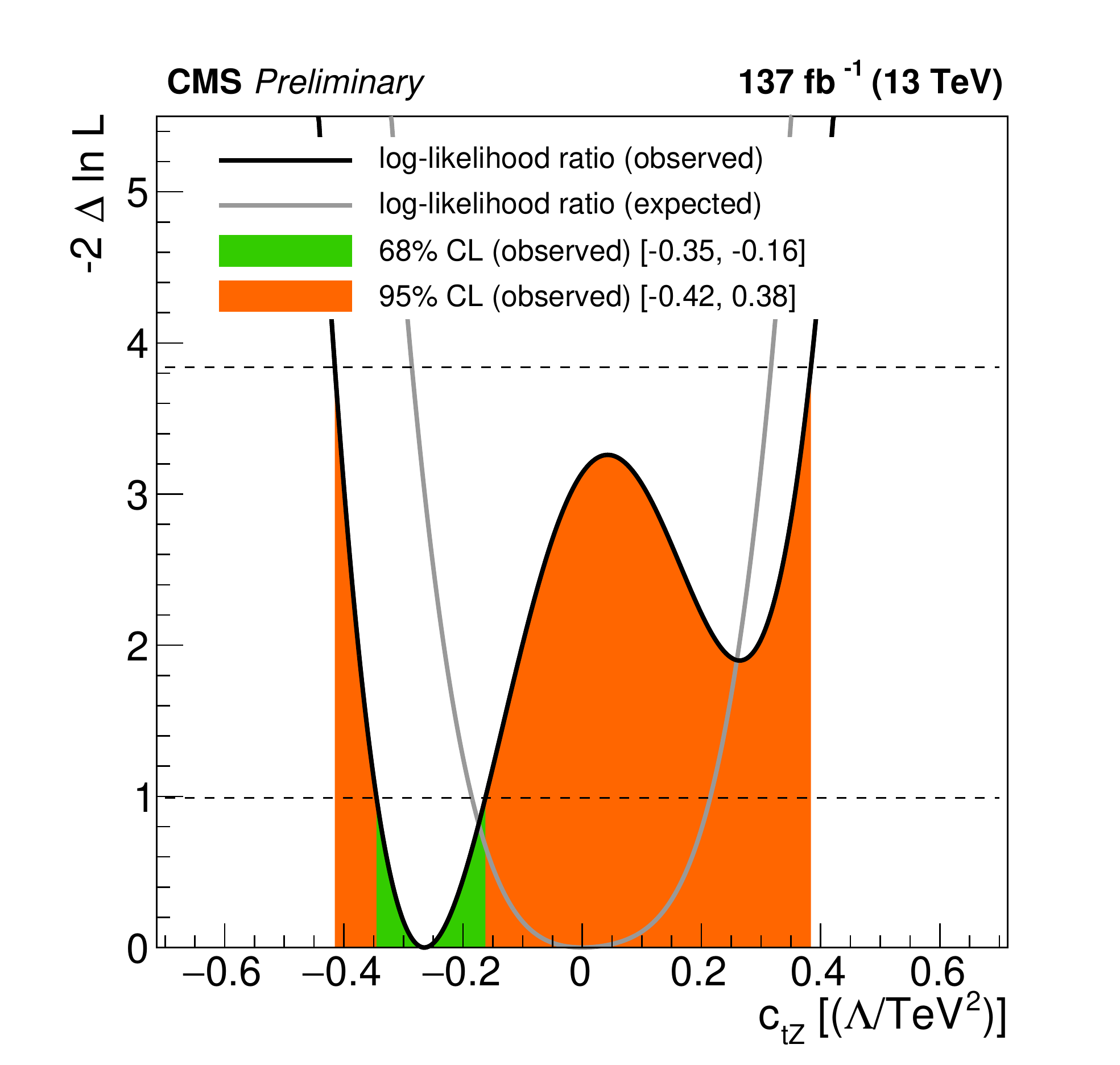}
    \includegraphics[width=6cm,height=6cm]{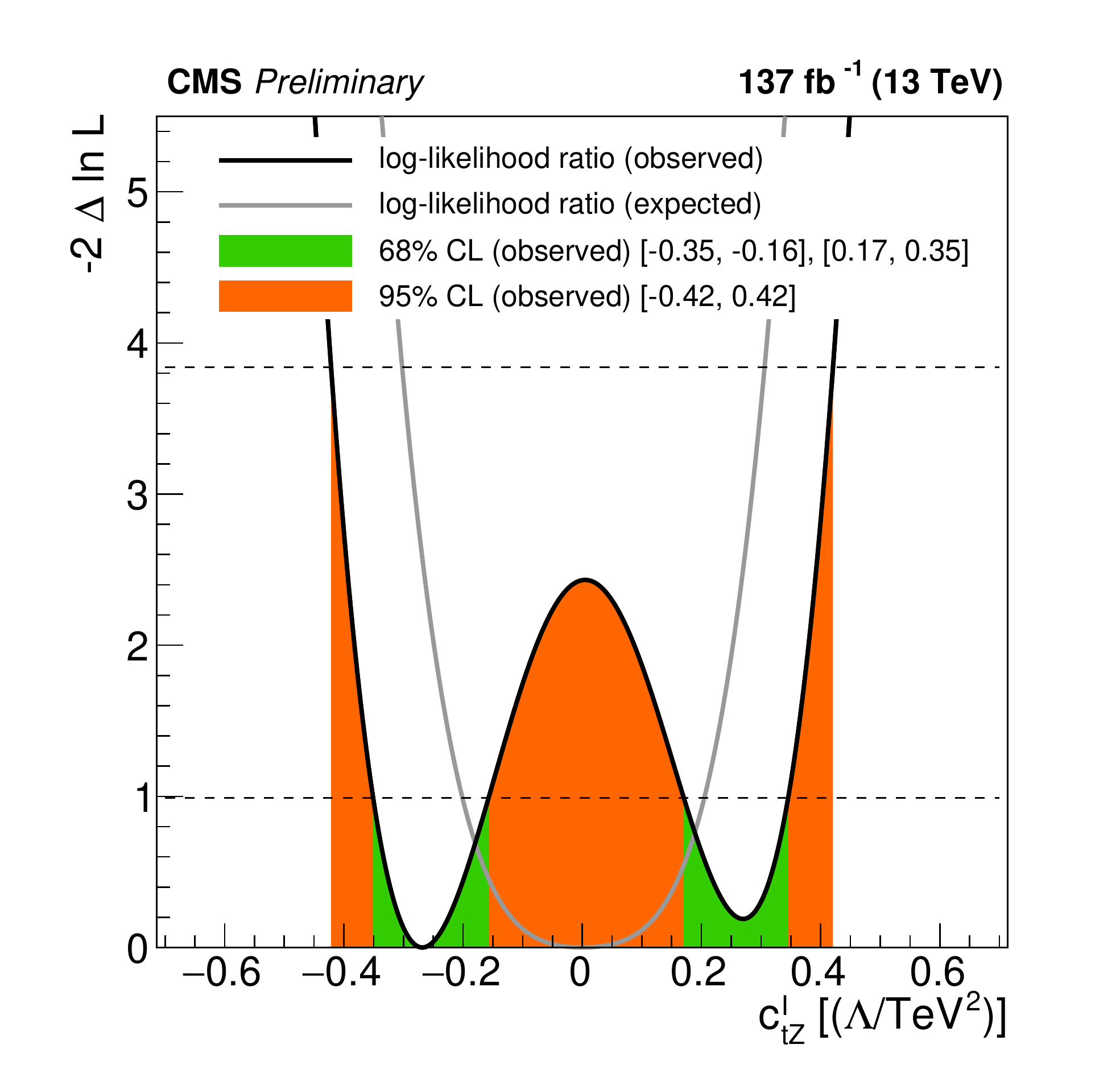}
    \includegraphics[width=6cm,height=6cm]{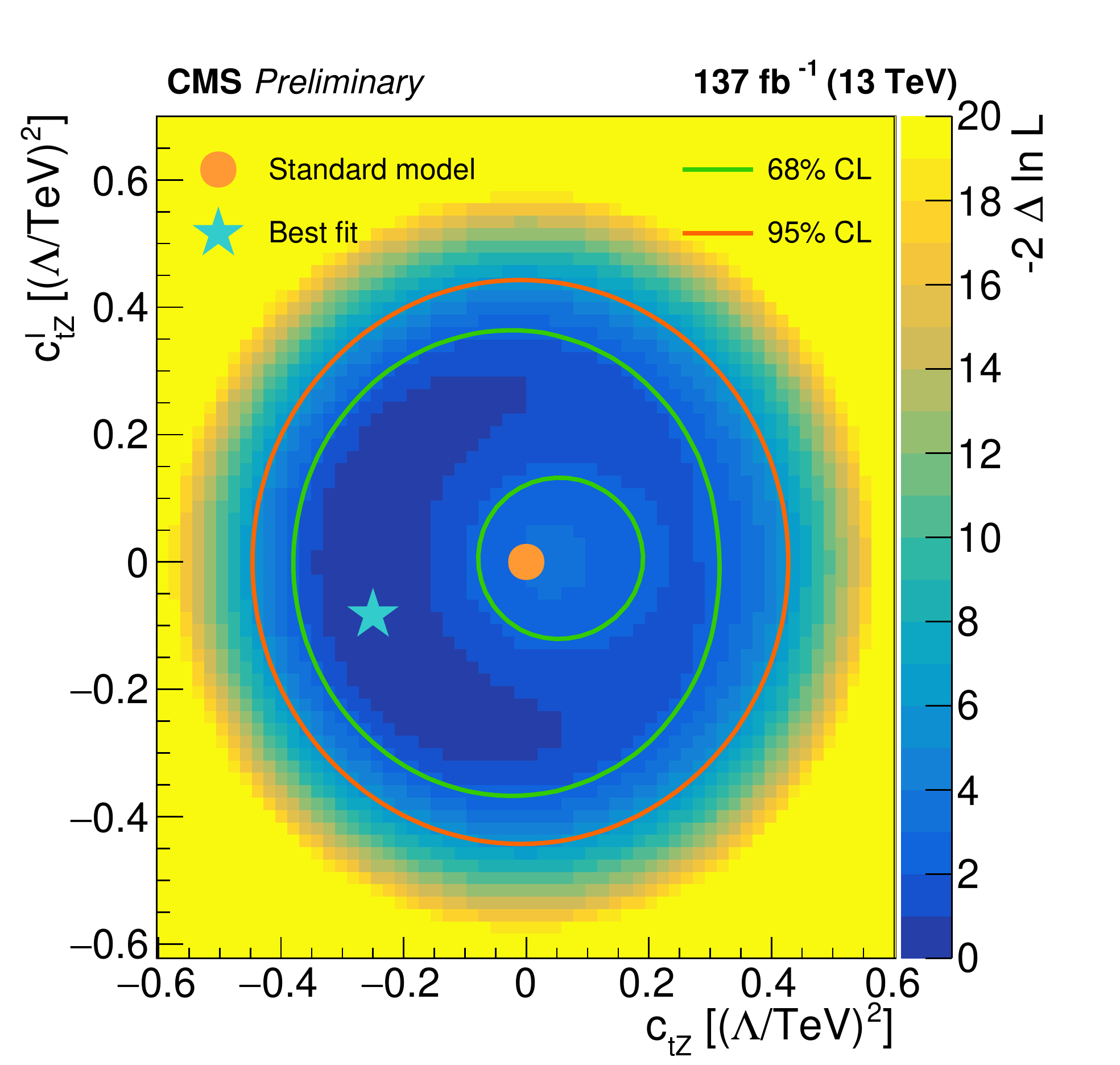}
    \caption{Best fit values for the explored EFT Wilson coefficients by CMS in the single$-$lepton channel~\cite{CMStty13tev_ljets}. Both the 1D and 2D scans are shown. \label{CMS-tty_EFT}}
\end{figure}

A maximum likelihood fit using the $\mathrm{p_T}(\gamma)$ spectrum, which is sensitive to such modifications, is performed to obtain $68\%$ and $95\%$ CL intervals on the targeted coefficients. The fit is performed in the signal regions only. The intervals for a given Wilson coefficient are obtained by either fixing the other WC to its SM value (1D), or simultaneously profiling the two WCs (2D). The results of both tests are shown in Figure \ref{CMS-tty_EFT}. No deviation from the SM values is observed. The 1D scans show more stringent intervals than the $\text{t}\bar{\text{t}}Z$ measurements. This is partially because models with non-zero WC values predict a harder $\mathrm{p_T}(\gamma)$ spectrum, which is not observed in the tails of the data distribution. The precision with which CMS can reconstruct photon kinematics is a major contributing factor to this measurements ability to improve upon the latest limits.
\vspace{-6pt}

\begin{figure}[H]
    \centering
    \captionsetup{justification=centering}
    \includegraphics[width=6cm,height=6cm]{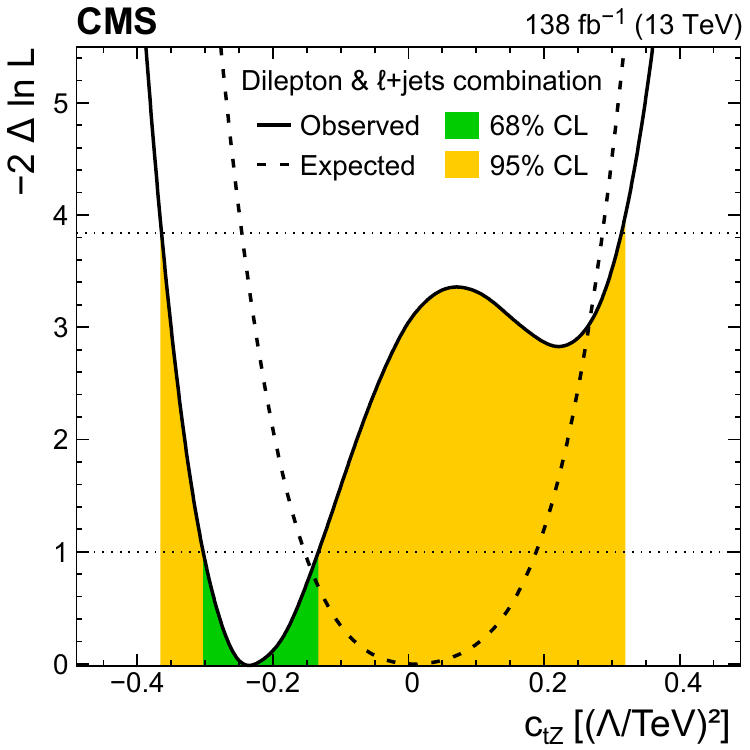}
    \includegraphics[width=6cm,height=6cm]{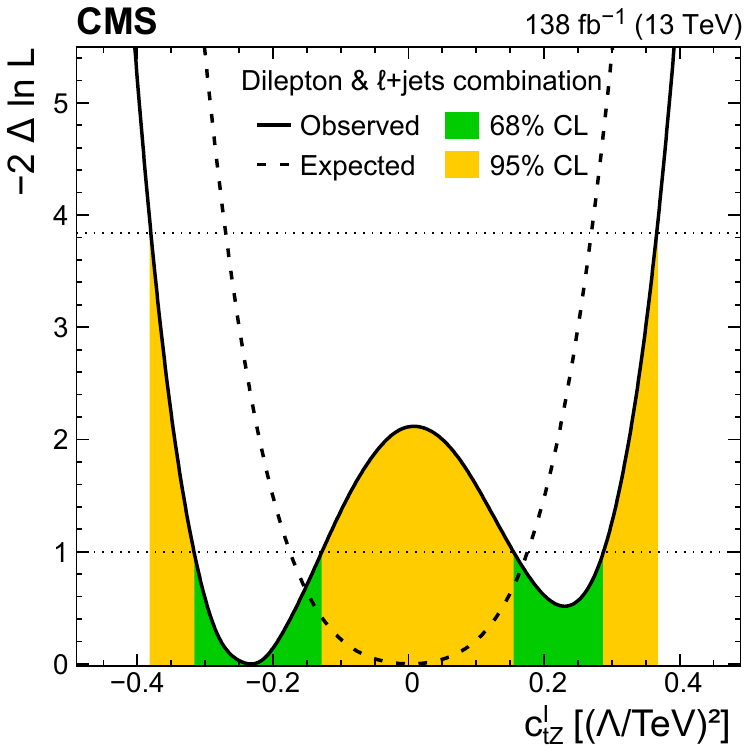}
  	\includegraphics[width=6cm,height=6cm]{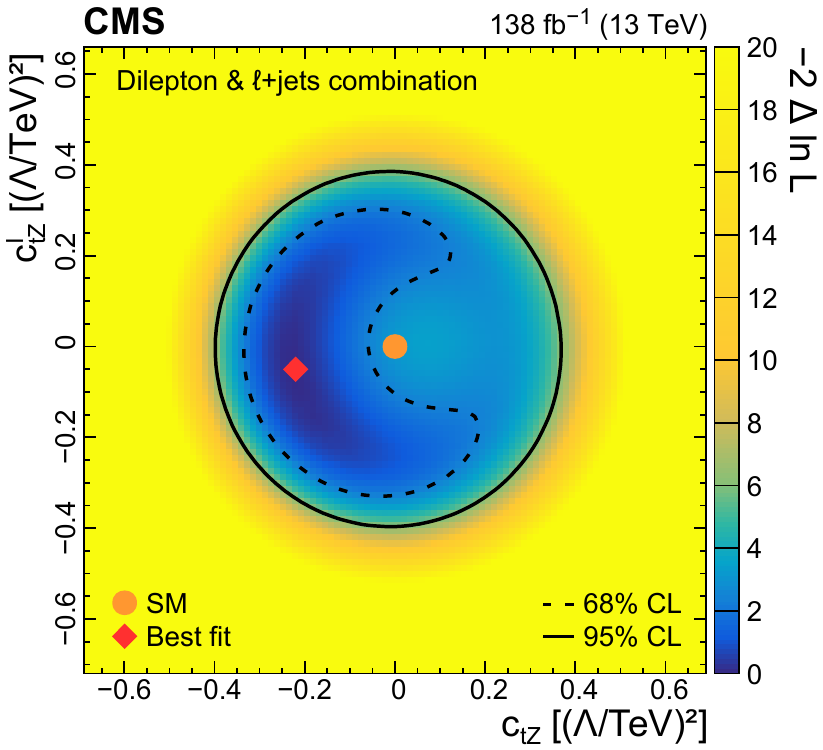}
    \caption{Distributions of the observed (solid line) and expected (dashed line) negative log-likelihood difference from the best fit value for the one-dimensional and two-dimensional scans of the studied Wilson coefficients. The results are obtained from the fit to data using the photon $p_T$ distribution. The plots shown here are from the combination of the single lepton and dilepton analyses~\cite{CMStty13tev_dilep}.\label{CMS-tty-EFT-scans}}
\end{figure}

\begin{figure}[H]
    \centering
    \captionsetup{justification=centering}
    \includegraphics[width=8.0cm,height=8.0cm]{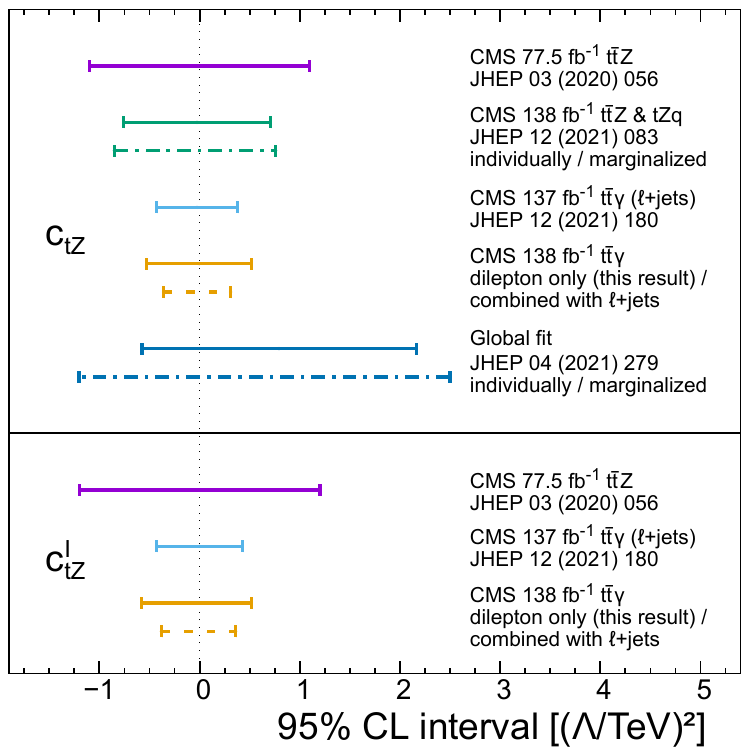}
    \caption{Comparison of observed 95\% CL intervals for the two Wilson coefficients, $c_{tZ}$ (\textbf{upper panel}) and $c_{tZI}$ (\textbf{lower panel}) from CMS measurements of: $\text{t}\bar{\text{t}}Z$, $\text{t}\bar{\text{t}}\gamma$ single lepton, $\text{t}\bar{\text{t}}\gamma$ dilepton. The results are shown from the one-dimensional scans, i.e., all other Wilson coefficients have values set to zero. The dashed lines indicate the results from the combination with the single$-$lepton channel. In the case of the global fit and the $\text{t}\bar{\text{t}}Z+tZq$, the solid lines represent the result where all Wilson coefficients are fixed to zero, whereas the dashed lines show the results from the marginalised limits. The tightest constraint to date on these Wilson coefficients comes from the combination of the $\text{t}\bar{\text{t}}\gamma$ single lepton and dilepton channels~\cite{CMStty13tev_dilep}. \label{CMS-tty-EFT-comparison}} 

\end{figure}

\textls[-5]{The CMS dilepton measurement performs a profile likelihood fit in the same way as the inclusive measurement, to obtain the best fit values for the Wilson coefficients probed. A combined profile likelihood fit is also performed with the single-lepton analysis. Although the dilepton channel benefits from a higher purity of signal, the single-lepton channel profits from a higher number of signal events with a high $p_T$ photon, making it sensitive to modifications in the kinematics of the photon caused by anomalous Wilson coefficient values. The 1D and 2D scans of the Wilson coefficients in both the dilepton and combined fits can be found in Figure \ref{CMS-tty-EFT-scans}. No sign of anomalous couplings is observed. A comparison with the constraints from other measurements is also shown in Figure \ref{CMS-tty-EFT-scans}. The results in this publication provide the best limits to date on the $c_{tZ}$ and $c_{tZ}^I$ Wilson coefficients in Figure \ref{CMS-tty-EFT-comparison}.}

\vspace{-6pt}

\section{Measurements at the HL-LHC and Future Colliders}
Cross-section measurements of rare $\text{t}\overline{\text{t}}+X$ process are incredibly useful probes of top-quark couplings to gauge bosons and are therefore a key ingredient to furthering our knowledge at the high energy frontier. Anomalous couplings are predicted by several BSM theories including composite Higgs models, models with extra dimensions and those predicting vector-like quarks~\cite{future_EW_couplings,top_window_to_compositeness,SM_EFT_fits}.

The LHC has already produced a sizeable sample of rare top-quark processes, which has been used to measure the cross-sections for several $\text{t}\overline{\text{t}}+X$ processes with an uncertainty that is considered to be on the cusp of what is commonly referred to as a `precision measurement'. The large dataset from the HL-LHC will cement these measurements in the precision regime and allow more precise probes of anomalous couplings affecting these amplitudes. Extrapolations of current measurements to future datasets and accelerators provide estimates of what might be achieved, help to establish physics goals and highlight the improvements required to achieve them.

Although the HL-LHC will provide a huge rare top dataset, enhancing the boosted regime in particular, it is also interesting to look towards machines planned even further in the future. Several machines fall into this category and are typically designed to push the precision frontier through the clean environment provided via lepton collisions, or the energy frontier through the high energies achievable at large circular hadron colliders. Results from future lepton colliders are particularly interesting in the context of this article as ultra-precise measurements of top-quark EW interactions will be achievable. Future colliders, of both hadrons and leptons, at or above the energy frontier ($\geq$10\,TeV) have the potential to improve the sensitivity of Standard Model EFT (SMEFT) fits to new physics, particularly to four-fermion operators for which there is a strong increase in sensitivity at higher energies.

To fully harness the power of precision measurements in a truly model-independent search for new physics, it is best to take a global approach to SMEFT fits~\cite{SMEFIT}. This requires a combination of the broadest dataset possible in a high-dimensional fit of many operators affecting several SM processes into account. Several of these operators are of particular interest given the scope of this article, namely operators affecting top EW couplings. So far, this article has only discussed measurements of SMEFT parameters using $\text{t}\overline{\text{t}}Z/\gamma$ however, operators can affect several SM processes in many different ways and hence a global fit of these operators using many processes can provide important constraints.

\begin{figure}[H]
    \centering
    \captionsetup{justification=centering}
    \includegraphics[width=12cm,height=8.0cm]{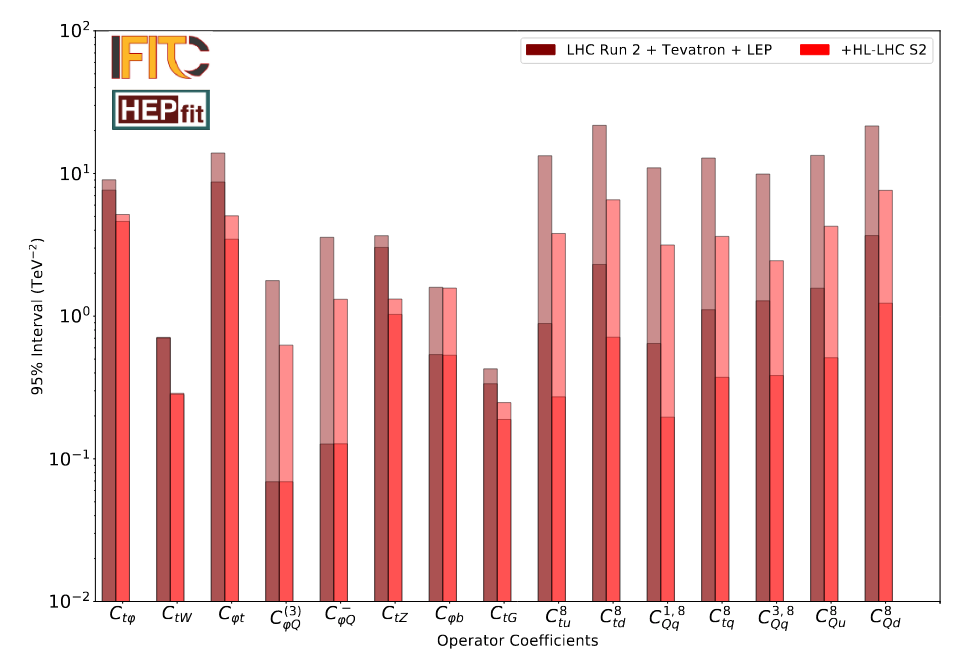}
    \caption{Comparison of expected 95\% confidence intervals on Wilson coefficients for dimension-six operators affecting top-quark production and decay measurements using the LHC Run 2 dataset and the HL-LHC dataset~\cite{snowmass_top}. Only linear terms proportional to $\Lambda^{-2}$ are accounted for in the dependence of the observables on the Wilson coefficients. The solid bars show the constraint of from the single parameter fits, while the translucent bars show the marginalised constraints from the global fit. \label{EFT_LHC_HLLHC}}
\end{figure}

The outlook for measurements of EFT parameters affecting top EW couplings has in fact been studied and reported in several publications. A comparison of the expected 95\% confidence interval for several EFT operators, using the LHC Run 2 dataset and the extrapolated values using the HL-LHC dataset are shown in Figure \ref{EFT_LHC_HLLHC}. The figure shows the results from a global EFT fit performed in Ref.~\cite{snowmass_top}. A full list of analyses included in the global fit can be found in Table \ref{EFT_combined_fit_processes}. It should be noted that, although the HL-LHC data are shown to bring an improvement to the global fits of almost all of the operators in question (Figure \ref{EFT_LHC_HLLHC}), the individual $95\%$ confidence intervals on operators $C_{\varphi Q}^{-}$ and $C_{\varphi Q}^{3}$ are not enhanced. This is due to their reliance on the legacy $e^+e^- \rightarrow \mathrm{b\overline{b}}$ measurements of $R_b$ and $A^{bb}_{FBLR}$ at the Z-pole from LEP and SLC. The inclusion of the Tevatron s-channel single-top measurement provides complementary constraining power on these operators and is still the most sensitive measurement of this process which, at the time of writing this article, still illudes measurements at the LHC.

Not all processes used in the global fit are relevant for this article; however, the plot highlights the importance of $\text{t}\bar{\text{t}}+X$ measurements at the present and in the future. All projections (including for lepton colliders discussed later) are based on similar approximations to the 'S2' scenario used in projections of Higgs boson measurements~\cite{S2_scenario} where many statistical and experimental uncertainties scale as $\frac{1}{\sqrt{L_{int}}}$, with $L_{int}$ representing the integrated luminosity. With respect to uncertainties at the end of the LHC Run 2, the complete HL-LHC program approximates that experimental uncertainties will reduce by a factor of 5, while theory and modelling uncertainties are reduced by a factor of two. The reduction in theory uncertainties assumes that $\mathrm{N^2LO}$ calculations will be achieved for the rare top processes and that large steps forward in Monte Carlo modelling are made in the next 10 years, ready for when the new colliders are expected to start running.

\begin{table}[H]
\captionsetup{justification=centering}
\caption{Measurements included in the top-quark EW sector EFT fit~\cite{snowmass_top}. The table includes the process, observable, centre-of-mass energy, integrated luminosity and experiment for each measurement. Where the experiment is cited as LHC, a combination of ATLAS and CMS measurements were used. Where Tevatron is cited, a combination of CDF and D0 results were used. LEP/SLD refers to different experiments from these two accelerators. 
\label{EFT_combined_fit_processes}}

\setlength{\tabcolsep}{4.3mm}
\resizebox{\textwidth}{!}{\begin{tabularx}{\textwidth}{cclll}
\toprule
\textbf{Process}	& \textbf{Observable} & \textbf{$\sqrt{s}$} & \textbf{Luminosity} & \textbf{Experiment} \\
\midrule
$\text{pp}\rightarrow \text{t}\overline{\text{t}}$		    & $\frac{d\sigma}{dm_{\text{t}\overline{\text{t}}}}$     & $13$\,TeV    & $140~\mathrm{fb^{-1}}$  & CMS \\
$\text{pp}\rightarrow \text{t}\overline{\text{t}}$           & $\frac{dA_C}{dm_{\text{t}\overline{\text{t}}}}$        & $13$\,TeV    & $140~\mathrm{fb^{-1}}$  & ATLAS \\
$\text{pp}\rightarrow \text{t}\overline{\text{t}}H+tHq$      & $\sigma$                                 & $13$\,TeV    & $140~\mathrm{fb^{-1}}$  & ATLAS \\
$\text{pp}\rightarrow \text{t}\overline{\text{t}}Z$          & $\frac{d\sigma}{dp_{T}^{Z}}$             & $13$\,TeV    & $140~\mathrm{fb^{-1}}$  & ATLAS \\
$\text{pp}\rightarrow \text{t}\overline{\text{t}}\gamma$     & $\frac{d\sigma}{dp_{T}^{\gamma}}$        & $13$\,TeV    & $140~\mathrm{fb^{-1}}$  & ATLAS \\
$\text{pp}\rightarrow tZq$                     & $\sigma$                                 & $13$\,TeV    & $77.4~\mathrm{fb^{-1}}$ & CMS \\
$\text{pp}\rightarrow \text{t}\gamma q$               & $\sigma$                                 & $13$\,TeV    & $36~\mathrm{fb^{-1}}$   & CMS \\
$\text{pp}\rightarrow \text{t}\overline{\text{t}}W$          & $\sigma$                                 & $13$\,TeV    & $36~\mathrm{fb^{-1}}$   & CMS \\
$\text{pp}\rightarrow \text{t}\overline{\text{b}}$ (s-chan)   & $\sigma$                                 & $8$\,TeV     & $20~\mathrm{fb^{-1}}$   & LHC \\
$\text{pp}\rightarrow tW$                      & $\sigma$                                 & $8$\,TeV     & $20~\mathrm{fb^{-1}}$   & LHC \\
$\text{pp}\rightarrow \text{tq}$ (t-chan)              & $\sigma$                                 & $8$\,TeV     & $20~\mathrm{fb^{-1}}$   & LHC \\
$\text{pp}\rightarrow Wb$                      & $F_0, F_L$                               & $8$\,TeV     & $20~\mathrm{fb^{-1}}$   & LHC \\
$\text{p}\overline{\text{p}}\rightarrow \text{t}\overline{\text{b}}$ (s-chan)& $\sigma$                         & $1.96$\,TeV  & $9.7~\mathrm{fb^{-1}}$  & Tevatron \\
$e^+e^-\rightarrow \text{b}\overline{\text{b}}$       & $R_b, A_{FBLR}^{bb}$                  & $91$\,GeV & $202.1~\mathrm{pb^{-1}}$& LEP/SLC \\
\bottomrule
\end{tabularx}}
\end{table}

This study highlights the need for further advances in theoretical calculations and modelling for HL-LHC measurements where, according to the current `S2' model for projections, theory uncertainties will for the first time dominate over experimental and statistical sources. The current state of the art for the theory predictions of the relevant processes, as well as the desires of the experimental community for future predictions are reported here.

The latest $\text{t}\bar{\text{t}}+W$ calculations have been discussed at length in the relevant section in this article as the area is particularly active. To summarise, the latest calculations have been performed using matrix element using perturbative calculations with precision up to the NLO terms in QCD and include additional next-to-next-to-leading log (NNLL) effects~\cite{ttV_NLO_NNLL}, as well as predictions for NLO+NNLL in QCD with NLO EW corrections. Full off-shell calculations up to NLO in QCD~\cite{ttW_NLO_offshell_1,ttW_NLO_offshell_2,ttW_NLO_offshell_3} have also recently been developed and now with the possibility to combine NLO EW and QCD corrections to off-shell $\text{t}\overline{\text{t}}W$~\cite{ttW_offshell_QCD_EW} and a procedure to apply the full off-shell corrections within the NLO+PS setup~\cite{ttW_offshell_NLO_PS}. Future NNLO calculations could bring a factor of two improvement in the precision of the~calculation.

NLO QCD calculations of $\text{t}\overline{\text{t}}\gamma$ have been available for a while~\cite{hardphoton_ttgamma}. The most recent NLO calculation was in fact for the process $\text{t}\overline{\text{t}}\gamma + tW\gamma$~\cite{tWy_precise} in the $\text{e}\mu$ final state. The inclusion of NNLO QCD corrections in a full $\text{t}\overline{\text{t}}\gamma$ calculation will become necessary if the full potential of the data at the HL-LHC is to be exploited.

The latest $\text{t}\overline{\text{t}}Z$ cross-section calculation is of NLO QCD+EW precision. This not only takes into account the $Z/\gamma$ interference, but also includes the off-shell $\text{t}\overline{\text{t}}\gamma^{\star}$ contributions. The theory uncertainty in this calculation is $^{+0.09}_{-0.10}$~\cite{ttZ_theoryXS_ATLAS_1,ttZ_theoryXS_2,ttZ_theoryXS_3}. This is mainly a result of the proton PDF, QCD scale and $\alpha_S$. The measurements in Section \ref{Section:ttZ} show that the total systematic uncertainty of the inclusive and differential cross-section measurement are already very close to this. A more precise theory calculation in the future would have a great impact on the achievable precision of future EFT measurements sensitive to effects from the $\mathcal{O}_{tZ}$~operator.

Figure \ref{EFT_LHC_HLLHC} shows the Wilson coefficients for several EFT operators along the x-axis. $\text{t}\bar{\text{t}}X$ processes are sensitive to the first six couplings from the left. The remaining couplings often affect top-pair production via QCD mechanisms and can be investigated more precisely using other $\text{t}\bar{\text{t}}$ processes. Differential measurements of $\text{t}\bar{\text{t}}Z$ and $\text{t}\bar{\text{t}}\gamma$ as a function of the Z boson or photon transverse momentum, respectively, are essential probes of the effects of the $\mathrm{O_{tZ}}$ operator. With increasing statistics, several measurements of rare top processes could be measured to much greater precision. Notably, precise differential measurements of $\text{t}\bar{\text{t}}W$ would provide essential information on this key background to measurements of $\text{t}\bar{\text{t}}H$ and four-top process in multi-lepton final states to name but a few.

Across all selected operators, a factor of two to four times the current Run 2 limits is expected with the HL-LHC dataset, both for the individual and marginalised bounds. The exceptions to this are the individual bounds of $C_{\varphi Q}^{-}$ and $C_{\varphi Q}^{3}$, which are very dependent on the bounds from the $\mathrm{Zb\overline{b}}$ measurements at the Z-pole. Sensitivity to operators affecting EW couplings could be dramatically improved in the future through the harvesting and analysis of data large datasets in the boosted regime~\cite{EFT_at_HE}. An additionally interesting insight from this reference is that, although not included in the fits performed in the document, the two-quark two-lepton ($\mathcal{O}_{qq\ell\ell}$) operators can be probed at the LHC and beyond and by including analyses targeting for instance the off-Z-peak dilepton invariant mass region in $\mathrm{\text{t}\overline{\text{t}}\ell^+\ell^-}$, the sensitivity of EFT fits can be enhanced.


Although the HL-LHC provides a much larger dataset with which to study EW couplings, the processes that provide the most sensitivity remain $\text{pp} \rightarrow \text{t}\bar{\text{t}}Z$ and $\text{pp} \rightarrow \text{t}\bar{\text{t}}\gamma$. Future lepton colliders provide an excellent opportunity to perform high-precision tests for anomalous EW couplings affecting top-quark pair processes. One of the benefits of $e^+e^-$ machines is that once the centre-of-mass energy exceeds twice the top mass, the dominant $\text{t}\bar{\text{t}}$ production mechanism becomes $ \mathrm{e^+e^-} \rightarrow \mathrm{Z/}\gamma \rightarrow \mathrm{\text{t}\overline{\text{t}}}$, providing direct access to the top-quark EW couplings in a very clean environment. Furthermore, lepton colliders can distinguish the coupling between the top quark and photon from the top-quark coupling with a Z boson. At circular lepton colliders, this is facilitated via a measurement of the final state polarisation in semileptonic top-quark decays, whereas at a linear collider this can be done using different beam polarisations configurations~\cite{topEFT_global_lepton_colliders,EW_vertices_ILC,precise_EW_couplings,CLIC_top}.

Figure \ref{EFT_HLLHC_LCs} compares expected limits on the different EFT operator coefficients using combinations of the data collected from the HL-LHC combined with data taken in the final stages of four different future lepton colliders: the CEPC, FCC, ILC  and CLIC~\cite{snowmass_top}. Important information on the different working configurations of each future machine is shown in Table \ref{future_collider_comparison}. Though not as important for the processes and operators discussed here, it is worth noting that the different runs have different centre-of-mass energies above the top-quark pair production threshold, which can be used to disentangle the four-fermion $e^+e^-\text{t}\overline{\text{t}}$ operator coefficients from the two-fermion operator coefficients. This is because the four-fermion operators scale quadratically with energy whereas the two-fermion operators either remain constant or grow linearly. Given the energies above the $\text{t}\bar{\text{t}}$ threshold in the circular collider scenarios are very close, this disentanglement is more difficult with such~machines.

\begin{table}[htbp] 
\newcolumntype{C}{>{\centering\arraybackslash}X}
\newcolumntype{b}{>{\hsize=.3\hsize}X}
\captionsetup{justification=centering}
\caption{Table of the working configurations for several future $e^+e^-$ colliders from Ref.~\cite{snowmass_top}. The machines listed in the table are: the International Linear Collider (ILC), the Circular Electron–Positron Collider (CEPC), the Compact Linear Collider (CLIC) and the Future Circular Lepton Collider (FCC-ee). The polarisation, energy and luminosity for 3 to 4 different running stages are listed along with references to the relevant documentation. \label{future_collider_comparison}}
\begin{tabularx}{\textwidth}{bCbb}
\toprule
\textbf{Machine}	& \textbf{Polarisation} & \textbf{Energy} & \textbf{Luminosity} \\
\midrule
ILC	\cite{ILC}	& P($e^+, e^-$): ($\pm30\%$,$\mp80\%$) & $250$ GeV & $2 ab^{-1}$  \\
        &                                      & $500$ GeV & $4 ab^{-1}$  \\
        &                                      & $1$ TeV   & $8 ab^{-1}$  \\
CLIC~\cite{CLIC}	& P($e^+, e^-$): ($\pm30\%$,$\mp80\%$) & $380$ GeV & $1 ab^{-1}$  \\
        &                                      & $1.4$ TeV & $2.5 ab^{-1}$  \\
        &                                      & $3$ TeV   & $5 ab^{-1}$    \\
FCC-ee~\cite{FCC_snowmass}  &\multirow{3}{4em}{Unpolarised}        & Z-pole  & $150 ab^{-1}$ \\
        &                                      & $240$ GeV & $5 ab^{-1}$   \\
        &                                      & $350$ GeV & $0.2 ab^{-1}$ \\
        &                                      & $365$ GeV & $1.5 ab^{-1}$ \\
CEPC~\cite{FCC_snowmass}	&\multirow{3}{4em}{Unpolarised}        & Z-pole  & $57.5 ab^{-1}$ \\
        &                                      & $240$ GeV & $20 ab^{-1}$  \\
        &                                      & $350$ GeV & $0.2 ab^{-1}$ \\
        &                                      & $360$ GeV & $1 ab^{-1}$   \\
\bottomrule
\end{tabularx}
\end{table}

The data from circular colliders (FCC-ee and CEPC) operating at centre-of-mass energies equal to and slightly above the $\text{t}\bar{\text{t}}$ threshold, are expected to improve in the constraints on the bottom and top operators at the HL-LHC by a factor of 2 to 5 for several two-fermion operators. The constraining power on four-fermion operators is limited by the energy reach. The data for the linear colliders (ILC and CLIC) was simulated at two centres of mass energies above the $\text{t}\bar{\text{t}}$ threshold and provides impressive constraints on all operators. As mentioned, it is due to these different collision energies that even the bounds on the four-fermion operators become competitive once the centre-of-mass energy surpasses $1$\,TeV.

\begin{figure}[H]
    \centering
    \captionsetup{justification=centering}
    \includegraphics[width=12cm,height=8.0cm]{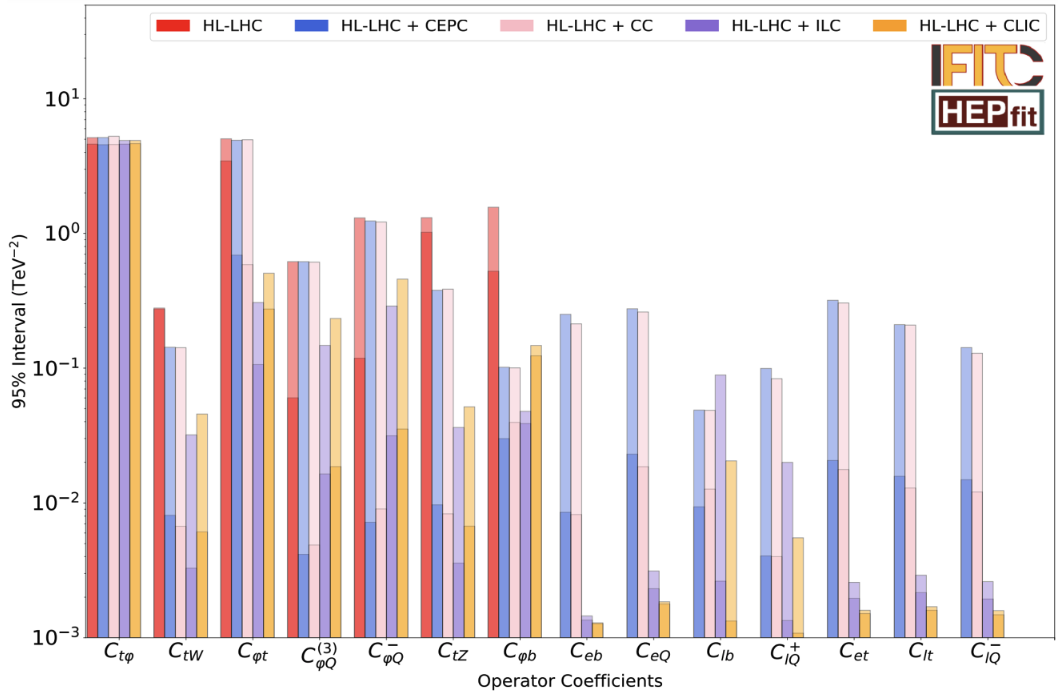}
    \caption{Comparison of expected 95\% confidence intervals combining data from the HL-LHC with data from several proposed lepton collider experiments~\cite{snowmass_top}. $q\overline{q}\text{t}\overline{\text{t}}$ and $C_{tG}$ coefficients are not shown in the figure as $e^+e^-$ collider measurements provide no additional sensitivity; however, all operators are included in the global fit. The solid bars show the constraint of from the single parameter fits, while the translucent bars show the marginalised constraints from the global fit. N.B. label HL-LHC$+$CC refers to the addition of FCC results. \label{EFT_HLLHC_LCs}}
\end{figure}

Looking further ahead, collisions at higher centre of mass (beyond 10TeV) could be achieved with for example a 100 km hadron collider, a linear electron–positron collider or compact circular muon collider~\cite{FCC_concept_design,ALEGRO_ES,Muon_collider}. As was alluded to earlier, the energy-growing sensitivity of the global SMEFT fits to new physics, especially through four-fermion operators, makes measurements at such machines invaluable. Given the absence of new physics signals, model-independent searches such as this provide one of the best chances of finding deviations from the SM and guiding the future of HEP.

\section{Conclusions}
The top quark is a unique particle in the known universe and while there are many priorities for high energy physics research, its distinctive features suggest it may have a special role in the SM. Therefore, understanding the top quark with absolute clarity remains a top priority for high energy physics experiments. The absence of new resonant particles has driven the development of novel methods to detect the presence of new physics, including indirect searches looking for anomalous couplings involving SM particles using EFT's. Such measurements require immense precision and a wealth of data. This has been the case for several years regarding the dominant QCD top-pair production mechanism. However, over the coming years, several rare top-pair processes will enter this regime, providing essential probes of anomalous couplings and new insights into where to look for this evasive new physics.

$\text{t}\bar{\text{t}}Z/\gamma$ measurements have had a sub $10\%$ precision for some time and have provided differential measurements and constraints on the relevant EFT operators while the most recent $\text{t}\bar{\text{t}}W$ measurements have a precision of around $7\%$, though unfortunately no differential or EFT measurement has been performed at the time of writing this article. With an influx of more data from Run 3 and beyond, the collected dataset of all the rare top-pair processes will be large enough to perform both differential and EFT measurements. Additionally, as we increase the dataset size the boosted regime will become more populated and, due to energy-growing effects in certain EFT operators, these regimes will become much more important and provide complementary constraints.

EFT measurements become so important going forward, allowing us to scrutinise the SM and use the power of precision measurement across diverse datasets to probe a wide range of operators in a model-independent manner to perform comprehensive searches for new physics. It is clear from the projections that as we look towards the HL-LHC, the achievable constraints on EFT parameters grow 2--4 times stronger in the top EW sector. However, these constraints will grow even stronger at future lepton colliders, which show further improvements of between a factor of 2--5.\vspace{6pt}

\funding{This research received no external funding.}

\institutionalreview{Not applicable.}

\informedconsent{Not applicable.}

\dataavailability{Not applicable.} 

\acknowledgments{Not applicable.}

\conflictsofinterest{The authors declare no conflict of interest.} 

\begin{adjustwidth}{-\extralength}{0cm}
\printendnotes[custom]
\reftitle{References}

\end{adjustwidth}
 \begin{adjustwidth}{-\extralength}{0cm}
\PublishersNote{}
\end{adjustwidth}
\end{document}